\documentclass[11pt,twoside]{article}
\usepackage{verbatim,epsfig}
\leftline{\copyright~ 1999 International Press}
\leftline{Adv. Theor. Math. Phys. {\bf 2} (1998) {\ 1105 -- 1140} }

\usepackage{verbatim}
\renewcommand{\theequation}{\thesection.\arabic{equation}}

\newif\ifnref

\nreffalse
\def\bfone{\relax{\rm 1\kern-.35em 1}}
\def\inbar{\vrule height1.5ex width.4pt depth0pt}
\def\IC{\relax\,\hbox{$\inbar\kern-.3em{\mss C}$}}
\def\ID{\relax{\rm I\kern-.18em D}}
\def\IF{\relax{\rm I\kern-.18em F}}
\def\IH{\relax{\rm I\kern-.18em H}}
\def\II{\relax{\rm I\kern-.17em I}}
\def\IN{\relax{\rm I\kern-.18em N}}
\def\IP{\relax{\rm I\kern-.18em P}}
\def\IQ{\relax\,\hbox{$\inbar\kern-.3em{\rm Q}$}}
\def\IR{\relax{\rm I\kern-.18em R}}
\font\cmss=cmss10
\font\cmsss=cmss9
\def\ZZ{\relax\ifmmode\mathchoice
{\hbox{\cmss Z\kern-.4em Z}}{\hbox{\cmss Z\kern-.4em Z}}
{\lower.9pt\hbox{\cmsss Z\kern-.4em Z}}
{\lower1.2pt\hbox{\cmsss Z\kern-.4em Z}}\else{\cmss Z\kern-.4em
Z}\fi}
\def\ZZs{\relax\ifmmode\mathchoice
{\hbox{\cmsss Z\kern-.4em Z}}{\hbox{\cmsss Z\kern-.4em Z}}
{\lower.9pt\hbox{\cmsss Z\kern-.4em Z}}
{\lower1.2pt\hbox{\cmsss Z\kern-.4em Z}}\else{\cmsss Z\kern-.4em
Z}\fi}
\def\a{\alpha} \def\b{\beta} 
 
\def\G{\Gamma} 
\def\L{\Lambda} 
\def\cA{{\cal A}} \def\cB{{\cal B}}
\def\cC{{\cal C}} 
\def\cE{{\cal E}}
\def\cF{{\cal F}} \def\cG{{\cal G}}

\def\cL{{\cal L}} 
\def\cN{{\cal N}} \def\cO{{\cal O}}
 \def\cQ{{\cal Q}}
\def\cR{{\cal R}} 
\def\nup#1({Nucl.\ Phys.\ $\us {B#1}$\ (}
\def\plt#1({Phys.\ Lett.\ $\us  {#1}$\ (}
\def\cmp#1({Comm.\ Math.\ Phys.\ $\us  {#1}$\ (}
\def\prp#1({Phys.\ Rep.\ $\us  {#1}$\ (}
\def\prl#1({Phys.\ Rev.\ Lett.\ $\us  {#1}$\ (}
\def\prv#1({Phys.\ Rev.\ $\us  {#1}$\ (}
\def\mpl#1({Mod.\ Phys.\ Let.\ $\us  {A#1}$\ (}
\def\ijmp#1({Int.\ J.\ Mod.\ Phys.\ $\us{A#1}$\ (}
\def\tit#1|{{\it #1},\ }
\def\ni{\noindent}
\def\tilde{\widetilde}
\def\bar{\overline}
\def\us#1{\underline{#1}}

\def\hat{\widehat}
\def\hyp{\vrule height 2.3pt width 2.5pt depth -1.5pt}

\def\Coeff#1#2{{#1\over #2}}
\def\Coe#1.#2.{{#1\over #2}}
\def\coeff#1#2{\relax{\textstyle {#1 \over #2}}\displaystyle}
\def\coe#1.#2.{\relax{\textstyle {#1 \over #2}}\displaystyle}

\def\shalf{\relax{\textstyle {1 \over 2}}\displaystyle}

\def\to{\rightarrow}
\def\notin{\hbox{{$\in$}\kern-.51em\hbox{/}}}
\def\shdot{\!\cdot\!}

\def\attac#1{\Bigl\vert
{\phantom{X}\atop{{\rm\scriptstyle #1}}\phantom{X}}}

\def\del{\partial}

\catcode`\@=12
\def\F{{F}}

\def\FF{{\cF}} % ?????
\def\GG{{\cal G}} % ?????
\def\taueff{\tau_{{\rm eff}}}
\def\PK{{P_{K3}}}

\def\h {{1\over 2}}

\def\ov {\overline}
\def\o {\over}
\def\Li {{\cal L}i}
\def\Uc {{\ov U}}

\def\th {\theta}

\def\Ga {\Gamma}

\def\tr {{\rm Tr}}
\def\det {{\rm det}}

\def\lf {\left}
\def\ri {\right}

\def\p {\partial}

\def\lf {\Big}
\def\ri {\Big}

\def\bp#1{B^{(#1)}}
\def\cp#1{C^{(#1)}}
\def\hp{\attac{{\rm {harmonic\atop part\hfill}}}}

\def\appB{A}

\def\eprt#1{{\tt #1}}
\def\nihil#1{{\sl #1}}
\def\br{\hfill\break}
\def\np {{ Nucl.\ Phys.} {\bf B}}
\def\pl {{ Phys.\ Lett.} {\bf B}}

\def\prd {{ Phys. Rev. D} }

\begin{document}

\begin{center}
\vspace{0.4in}
{\huge \bf Prepotential, Mirror Map \\[5mm]and $F$-Theory on $K$3}
\vspace{0.4in}

{\bf W. Lerche  , S. Stieberger}
\vspace{0.2in}

Institute for Theoretical Physics,\\
University of California,\\
Santa Barbara, CA 93106, USA

\vspace{0.1in}
and 
\vspace{0.1in}

CERN,CH-1211\\
Geneva 23, Switzerland
\vspace{0.2in}
\begin{abstract}
We compute certain one-loop corrections to $F^4$ couplings of the
heterotic string compactified on $T^2$, and show that they can be
characterized by holomorphic prepotentials $\GG$.  We then discuss how
some of these couplings can be obtained in $F$-theory, or more
precisely from 7--brane geometry in type IIB language. We in
particular study theories with $E_8\times E_8$ and $SO(8)^4$ gauge
symmetry, on certain one-dimensional sub-spaces of the moduli space
that correspond to constant IIB coupling.  For these theories, the
relevant geometry can be mapped to Riemann surfaces.  Physically, the
computations amount to non-trivial tests of the basic $F$-theory --
heterotic duality in eight dimensions.  Mathematically, they mean to
associate holomorphic 5-point couplings of the form ${\del_t}^5\GG\sim
\sum g_\ell \ell^5 {q^\ell\over 1-q^\ell}$ to $K3$ surfaces. This can
be seen as a novel manifestation of the mirror map, acting here between
open and closed string sectors.
\end{abstract}
\renewcommand{\thefootnote}{}
\footnotetext{\ e-print archive: {\texttt http://xxx.lanl.gov/abs/hep-th/9804176}}
\renewcommand{\thefootnote}{\arabic{footnote}}

\end{center}
\newpage

\setcounter{page}{1106} 

\pagenumbering{arabic}

\pagestyle{myheadings}
\markboth{\it PREPOTENTIAL,MIRROR MAP AND $F$-THEORY ON $K$3}{\it W. LERCHE, S. STIEBERGER}
\setcounter{page}{1106}
\section{Introduction}
\setcounter{equation}{0}
To support the various string duality conjectures, numerous tests have
been successfully performed. So far lacking is a more quantitative test
of the basic heterotic-$F$ theory \cite{Fth} duality in eight dimensions,
i.e., a comparison of non-trivial terms in the effective
action.

The terms we have in mind are certain corrections to the $\F^4$
couplings. These couplings are related to the special class of
``holomorphic'' \cite{WL} or ``BPS-saturated'' \cite{BK,elias}
amplitudes, which involve $n=1,2,4$ external gauge bosons (in theories
with 4,8 or 16 supercharges, resp.), and which are directly related to
the heterotic elliptic genus $\hat\cA_{2n+2}$ \cite{ellg} in $2n+2$
dimensions. It turns out that supersymmetry relates parity even
($i\xi\,\F^n$) and parity odd (${\theta\over2\pi} \F \wedge \F\wedge ..
\F$) sectors, and therefore one can conveniently combine the
theta-angle and the coupling constant $\xi$ into one complex coupling,
$\taueff$. Specifically, functionally non-trivial one-loop corrections
arise in $T^2$ compactifications of $2n+2$ dimensional $N=1$
supersymmetric heterotic strings generically as follows:
\begin{eqnarray}\label{ellgen}%\nonumber\\\
{\rm Re}[\taueff(T,U)]\, \F\wedge^n \F\ \, \sim\ \int {d^2\tau\over {\tau_2}}
Z_{(2,2)}(T,U,q,\ov q)\hat\cA_{2n+2}(\F,q) \attac{2n-form}%\\\nonumber
%\eqn\ellgen
\end{eqnarray}
Here, $T,U$ are the K\"ahler and complex structure moduli and
$Z_{(2,2)}$
is the partition function of the two-torus $T^2$ (we have switched off
any
Wilson lines inFrom analyticity, the imaginary part of
$\taueff$ is determined essentially by the same expression.\footnote{The
mild non-harmonicity that arises from massless states can easily be
kept under control.} For gauge theories with the indicated number
of supersymmetries, it is known that perturbative contributions arise
to one loop order only and have the form this formula)
\begin{eqnarray}\label{lntau}
\taueff(a)\ \equiv\ i\xi(a)+\Coeff1{2\pi}\theta(a)\ =\
\Coeff1{2\pi i}\ln[a/\Lambda]\ ,
%\eqn\lntau
\end{eqnarray}
where $a$ is the Higgs field and $\L$ a cutoff scale; in string theory,
there will be additional $\a'$ corrections. As is by now familiar,
logarithmic monodromy shifts correspond to physically irrelevant shifts
of the $\theta$--angle.

In heterotic compactifications on $K3\times T^2$ to four dimensions,
such couplings multiplying $\F^2$ are given by two derivatives of a
holomorphic prepotential, $\taueff(t)=(\del_t)^2\FF(t)$. They have been
computed for a variety of heterotic compactifications
\cite{enhancements,HM,fs} and reproduced in the dual type
II string setup \cite{kv,NexTwoDual}. Here one makes use of mirror
symmetry \cite{mirror} of suitable CY manifolds, to compute three-point
couplings of the generic form $\FF_{ttt} \equiv {\del_t}^3\FF= \sum
c_\ell \ell^3 {q^\ell\over 1-q^\ell}$, which in turn are equal to
$\del_t(\taueff)(t)$.

Analogous $\F^4$ couplings in $d=8$ have recently been analyzed in
detail \cite{BK,BFKOV,elias,ko}, focusing on heterotic and type I
string compactifications on $T^2$ and on gauge fields $\F$ belonging to
$E_8\times E_8$ or $SO(32)$. It was argued in \cite{fs} and shown in \cite{ko}
that such four-point couplings can be expressed in terms of holomorphic
 prepotentials, similar as in four dimensions.   This hints at the
existence of a superspace representation of the effective lagrangians;
however, to our knowledge, neither this formalism nor an analog of
``special geometry'' seems to have been developed.

In contrast to the couplings in $d=4$, the heterotic couplings
$\taueff$ in $d=8$ are supposedly exact at one loop order \cite{BFKOV}.
Indeed, for theories with 16 supercharges, the dilaton sits in a
gravitational multiplet and thus cannot mix with the vector moduli
space. Therefore there is no room for non-perturbative corrections, and
correspondingly the cutoff in (\ref{lntau}) is given by $\L=(\a')^{-1/2}$.
The couplings $\taueff$ thus provide an ideal, quantitative testing
ground for the heterotic-$F$ theory duality.

In the present paper, we will concentrate on certain couplings
$\taueff(T,U)$ arising in theories with $E_8\times E_8$ and $SO(8)^4$
gauge symmetry. In section 2, we will compute these couplings from the
heterotic string, which amounts to evaluating certain integrals that
were not treated before in the literature; some more technical material
is deferred to Appendices.

We will find that the couplings in the $T,U$ sub-sector derive from an
underlying prepotential $\cG$,
\begin{eqnarray}
(\taueff)_{ijkl}(T,U)\hp\!\! \equiv\
\GG_{ijkl}(T,U)\ &=&\ \del_i\del_j\del_k\del_l \GG(T,U)\ ,\\ 
&&i,j,...\in\{T,U\}\ ,\nonumber
\label{tauGG}
\end{eqnarray}

\noindent and that they formally have a structure that is very similar to the
well-known couplings in $4d$. That is, taking one more derivative, we
have
\begin{eqnarray}
\GG_{ijklm}(T,U)\ \equiv\
%\del_i (\taueff)_{jklm}(T,U) =
{\rm const}+\!\! \sum_{n_T, n_U}\!
 g(n_T\shdot n_U) n_in_jn_kn_ln_m
{{q_T}^{n_T}{q_U}^{n_U}\over 1-{q_T}^{n_T}{q_U}^{n_U}}\ ,
\label{counting}
\end{eqnarray}
\noindent where $q_T\equiv e^{2\pi iT}$ etc., and where $g$ are certain
integer coefficients. From the heterotic string point of view, this is
perhaps not too surprising, given that the structure of the modular
integrals is similar in spirit (though much more complicated) as in
four
dimensions.

The situation is however quite non-trivial when considered from a
``dual'' viewpoint, which is  $F$-theory \cite{Fth}. It is well-known that
heterotic compactifications on $T^2$ are dual to $F$-theory on $K3$, so
it is natural to expect that the above couplings should be computable
entirely from $K3$ data, at a purely geometrical level. However, it is
a priori not obvious how such holomorphic, quintic couplings could be
canonically related to $K3$.\footnote{Related tree-level
four-point functions for type IIA
strings compactified on $K3$ have been considered in \cite{HOCV},
but these couplings do not seem to be obviously
holomorphic; see also \cite{otherKthree}.}

Indeed, for a complex $d$-fold, the natural holomorphic couplings
correspond to intersections of $d$ two-forms, and this is why one can
use mirror symmetry on 3-fold CY's to compute the three-point couplings
$\FF_{ttt}$. For $K3$ the natural holomorphic couplings are two-point
functions (which are trivial, ie., constant \cite{MNKS,PADMKthree}),
and not five-point functions $\GG_{ttttt}$.  The question thus arises
what the geometrical principles are that lead to five-point functions
of the requisite structure, and in particular what the meaning of the
integral coefficients is. For three-point functions on 3-folds, the
coefficients are known to count rational curves. What they precisely
count in eq. (\ref{counting}) for a $K3$, may turn out to be an
interesting mathematical question; see section 4 for a few remarks
on curve counting on $K3$.

We have thus all reason to expect a novel mechanism of mirror symmetry
to be at work here, and it is the purpose of the paper to make some
first modest steps in uncovering it.  We will not find a complete
solution of the problem, but will at least show how to relate some
couplings to the geometry of $K3$.  This will be addressed in section
3, starting with a discussion of the 7--brane geometry for one of the
two models that we consider. We then show how the problem can be
simplified, by going to a subspace of the moduli space that corresponds
to constant IIB coupling and to finite order monodromies in the
$z$--plane. This allows to lift the geometry to Riemann surfaces, which
happen to be nothing but well-known Seiberg-Witten curves. In this way
we can find Greens functions with the requisite global properties, with
which we are able to compute some $F^4$ gauge couplings from geometry.
For the couplings of the $SO(8)^4$ model, we find perfect agreement
with the corresponding heterotic string results. We will conclude with
some remarks in section 4.

We intend to give a comprehensive treatment in a more detailed
forthcoming analysis.

\section{Heterotic Computation of Holomorphic Prepotentials in $d=8$}
\setcounter{equation}{0}

Compactifications of the ten--dimensional heterotic string  on a torus
have in common $s$ N=1 abelian vector multiplets, whose scalars are the
$T$ and $U$ moduli of the torus $T^2$, besides $s-2=0,\ldots,16$ Wilson
lines; for some canonical examples, see Table 1.   The bosonic content
of the supergravity multiplet in $d=8$ contains two more vectors and
one real scalar, which is the dilaton, in addition  to the graviton and
an anti--symmetric  tensor field. The manifold of the scalar fields is
described by the K\"ahlerian coset space  $SO(s,2)/[SO(s)\times
SO(2)]\times SO(1,1)$. The tree--level action for N=1, $d=8$
supergravity has been constructed up to the two--derivative level in
\cite{sugra}  and may be obtained by performing a  dimensional reduction of
the ten--dimensional heterotic string action.  Due to the number of
fermionic zero modes, string one--loop corrections start at the
four--derivative level. In the following we focus on the
(four--derivative) couplings of the scalars  of the vector multiplets,
and derive their underlying prepotential.

{\vbox{{
$$
\vbox{\offinterlineskip\tabskip=0pt
\halign{\strut\vrule#
%%%%%%%%%%%%%%%%%%
&~$#$~\hfil
&\vrule#
&~$#$~\hfil
&~$#$~\hfil
&~$#$~\hfil
&\vrule#
%&~$#$~\hfil
%&~$#$~\hfil
%&~$#$~\hfil
%&\vrule#
%&~$#$~\hfil
%&~$#$~\hfil
%&\vrule#
%&~$#$~\hfil
%&\vrule#
%&~$#$~\hfil
%&\vrule#
\cr
%%%%%%%%%%%%%%%%%%
\noalign{\hrule}
&
{\rm gauge\ group}
&&
{\rm max.\ pert.\ enhanced}
&&
{\rm sublocus}
&
\cr
%&&
%{\rm modular\ group}
%&&
%{\rm shift}
%&
%\cr
%%%%%%%%%%%%%%%%%%
\noalign{\hrule}
&
E_8^2\times U(1)^2
&&
E_8\times E_8\times SU(3)
&&
U=T=\rho
%&&
%\Ga_T\times\Ga_U
%&&
%1
&
\cr
%%%%%%%%%%%%%%%%%%
%\noalign{\hrule}
&
E_7^2\times SU(2)^2\times U(1)^2
&&
E_7\times E_7\times SO(8)
&&
U=-1/T=i+1
%&&
%\Ga_0(2)_T\times\Ga^0(2)_U
%&&
%\ZZ_2
&
\cr
%%%%%%%%%%%%%%%%%%
%\noalign{\hrule}
&
E_6^2\times SU(3)^2\times U(1)^2
&&
E_6\times E_6\times E_6
&&
U=-1/T=\rho-1
%&&
%\Ga_0(3)_T \times\Ga^0(3)_U
%&&
%\ZZ_3
&
\cr
%%%%%%%%%%%%%%%%%%
%\noalign{\hrule}
&
SO(8)^4\times U(1)^2
&&
SO(8)^4\times U(1)^2
&&
-
%&&
%\Ga_0(4)_T\times\Ga_U
%&&
%\ZZ_2\times\ZZ_2
&
\cr
%%%%%%%%%%%%%%%%%%
\noalign{\hrule}}
\hrule}$$
\vskip10pt}}}

{\vbox{{
$
\vbox{\offinterlineskip\tabskip=0pt
\halign{\strut\vrule#
&~$#$~\hfil
&~$#$~\hfil
&~$#$~\hfil
&\vrule#
\cr
%%%%%%%%%%%%%%%%%%%
\noalign{\hrule}
&
{\rm modular\ group}
&&
{\rm shift}
&
\cr
%%%%%%%%%%%%%%%%%%
\noalign{\hrule}
&
\Ga_T\times\Ga_U
&&
1
&
\cr

%%%%%%%%%%%%%%%%%%
%%\noalign{\hrule}
&
\Ga_0(2)_T\times\Ga^0(2)_U
&&
\ZZ_2
&
\cr
%%%%%%%%%%%%%%%%%%
%\noalign{\hrule}
&
\Ga_0(3)_T \times\Ga^0(3)_U
&&
\ZZ_3
&
\cr
%%%%%%%%%%%%%%%%%%
%\noalign{\hrule}
&
\Ga_0(4)_T\times\Ga_U
&&
\ZZ_2\times\ZZ_2
&
\cr
%%%%%%%%%%%%%%%%%%
\noalign{\hrule}}
\hrule}$$
$

\noindent{\bf Table 1:}
{\sl Examples of heterotic gauge groups, possible maximal gauge
symmetry enhancements, and their shifts in the Narain lattice
$SO(18,2)$.  We also listed the duality groups in the $T,U$ subsector
that we find to be left unbroken by the Wilson lines.}
\vskip10pt}}}

In the following, we will mainly consider the first and the last of
these examples, namely with A: $E_8\times E_8$ and B: $SO(8)^4$
unbroken gauge symmetries.
\subsection{Example A: $E_8\times E_8$ Gauge Symmetry}
The generic form of the $(T,U)$--dependent couplings has been sketched
in eq.(\ref{ellgen}). It directly applies to couplings $\taueff$ multiplying
$E_8$ gauge fields, where Wilson lines are switched off; such couplings
have been computed in \cite{ko}. We will consider here the
couplings that are intrinsic to the $T^2$ sub-sector of the theory,
ie., the ones that multiply powers of the gauge fields $\F_T$, $\F_U$.
They are given by the following world--sheet modular integrals:
\begin{eqnarray}
\Delta_{\F_T^4}&=&{(U-\ov U)^2\o (T-\ov T)^2}
\int {d^2 \tau \o \tau_2} \sum_{(P_L,P_R)}\ \ov P_R^4\
q^{\h |P_L|^2}
\ov q^{\h |P_R|^2}\ {\bar E_4^2\o \ov \eta^{24}}\\
\Delta_{\F^2_T \F_U^2}&=&
\int {d^2 \tau \o \tau_2}
\sum_{(P_L,P_R)}\ \lf[\lf(|P_R|^2-
{1\o \pi \tau_2}\ri)^2-{1\o 2\pi^2\tau_2^2}\ri]\ q^{\h |P_L|^2}
\ \ov q^{\h |P_R|^2}\
\ {\bar E_4^2\o \ov \eta^{24}}\nonumber \\
\Delta_{\F_T^3 \F_U}&=&
{U-\ov U \o T-\ov T}\int {d^2 \tau \o \tau_2} \sum_{(P_L,P_R)}\
\lf(\ov P_R^3P_R-{3 \o 2\pi\tau_2} \ov P_R^2\ri)
\ q^{\h |P_L|^2}\ \ov q^{\h |P_R|^2}\ {\bar E_4^2\o \ov \eta^{24}}\
,\nonumber
\label{intFV}
\end{eqnarray}
\noindent with $q=e^{2\pi i\tau},\ov q=e^{-2\pi i\ov\tau}$ and the (complex)
Narain
momenta :
\begin{eqnarray}
P_L&=&{1\o \sqrt{2T_2 U_2}}(m_1+m_2\bar U+n_1\bar T+n_2\bar T\bar U)\nonumber \\
P_R&=&{1\o \sqrt{2T_2 U_2}}(m_1+m_2\bar U+n_1 T+n_2 T\bar U)\ \nonumber 
\end{eqnarray}
The above integrals\footnote{In \cite{ko} four--point couplings  involving the
$E_8$ gauge bosons and the graviton were considered.  Their
corresponding $\tau$--integrals show a quite different structure and
are expressed in terms of three prepotentials.  This is similar as for
$N=2$, $d=4$ theories where all two--point couplings involving  the
$E_8$ gauge bosons and the graviton  may be expressed by two functions
\cite{HM}.},  can be handled by using techniques developed in general in
\cite{DKLII,HM} and specifically in \cite{fs}. The corrections may be
understood as arising from the pieces ${{\hat E}_2^2E_4^2 \o \eta^{24}}
(\tr {\cal R}^2)^2$ and ${E_4^3 \o \eta^{24}} \tr ({\cal R}^4)$ of the
elliptic genus in ten dimensions \cite{ellg}, supplemented with the
corresponding zero modes $\ov P_R$; for more details of a similar
situation, see \cite{fs}. A direct string amplitude computation will be
presented in \cite{fsz} in the light of heterotic--type I string duality. As
result of lengthy calculations, we find

\begin{eqnarray}
\Delta_{\F_T^4}&=&16\pi i\lf(\p_T+{2\o T-\ov T}\ri)\p_T\lf(\p_T-{2\o
T-\ov T}\ri)
\lf(\p_T-{4\o T-\ov T}\ri)\GG\nonumber \\
&& -16\pi i{(U-\ov U)^4 \o (T-\ov T)^4}
\lf(\p_\Uc-{2\o U-\ov U}\ri)\p_\Uc\lf(\p_\Uc+{2\o U-\ov U}\ri)\nonumber \\
&&\times \lf(\p_\Uc+{4\o U-\ov U}\ri)\ov \GG\nonumber\\
\Delta_{\F^2_T \F_U^2}&=&16\pi i\lf(\p_U-{2\o U-\ov U}\ri)\lf(\p_U-{4\o U-\ov U}\ri)
\lf(\p_T-{2\o T-\ov T}\ri)\nonumber \\
&& \times \lf(\p_T-{4\o T-\ov T}\ri)\cG+hc.\\
\Delta_{\F_T^3 \F_U} 
&=&16\pi i\p_T\lf(\p_T-{2\o T-\ov T}\ri)\lf(\p_T-{4\o T-\ov T}\ri)
\lf(\p_U-{4\o U-\ov U}\ri)\cG\nonumber \\
&& -16\pi i{(U-\ov U)^2\o (T-\ov T)^2}\p_{\ov U}\lf(\p_{\ov U}+{2\o
U-\ov U}\ri)
\lf(\p_{\ov U}+{4\o U-\ov U}\ri)\nonumber \\&&\times \lf(\p_{\ov T}+{4\o T-\ov T}\ri)\ov \cG
\ .\nonumber
\label{DeltaFV}
\end{eqnarray}

It seems remarkable, at least formally, that these couplings integrate
to one and the same holomorphic prepotential $\GG$.
In the chamber $T_2>U_2$, it is given by
\begin{eqnarray}
\GG(T,U)&=&-{1\o 120}U^5
-{i\o (2\pi)^5}\sum_{(k,l)>0} g(kl)\ \Li_5 \big[
%e^{2\pi i(kT+lU)}
{q_T}^k {q_U}^l
\big]+\cQ(T,U)\ \nonumber \\ 
&& -{ic(0)\zeta(5)\o 64\pi^5},
\label{prepeight}
\end{eqnarray}
where the polylogarithm is defined by $(a \geq 1)$:
\begin{eqnarray}
\Li_a(z)=\sum_{p>0} {z^p \o p^a}\ , \qquad
{\rm with}\ \Big(z{\del\over\del z}\Big)^a\Li_a(z)={z\over 1-z}\ .
\label{LIS}
\end{eqnarray}
\noindent The sum in (\ref{prepeight}) runs over the positive roots
$k>0,\ l\in \ZZ\ \ \wedge\ \ k=0,\ l>0$.
The coefficients $g(n)$ are closely related to the chiral
light--cone partition function in the Ramond--sector,
${\cal C}(q,\ov q)=\ov \eta^{-24}\ {\rm Tr}_R[(-1)^F
q^{L_0-{c\o 24}}\ov q^{\ov L_0-{\ov c\o 24}}]$,
which for heterotic compactifications
is given by ${\cal C}(q,\ov q)=\ Z_{(2,2)}(q,\ov q)\
{\bar E_4^2 \o {\bar \eta^{24}}}$ with $Z_{(2,2)}(q,\ov
q)=\sum\limits_{(P_L,P_R)}
q^{\h|P_L|^2}\ov q^{\h |P_R|^2}$. Explicitly, they are
given by
\begin{eqnarray}
%{\cal A}(q,\ov q)&=2i\ Z_{2,2}(\tau,\ov\tau)\
%{\bar E_4^2 \o \ov \eta^{24}}\ ,\cr
{E_4^2\o \eta^{24}}&\equiv:&\sum_{n\geq -1} g(n)\ q^n.
%=\ov q^{-1}+504+73764 \ov q+2695040 \ov q^2+\ldots
\label{index}
\end{eqnarray}
The prepotential $\GG$ has modular weights $(w_T,w_U)=(-4,-4)$ under
$T$-- and $U$--duality. It is determined up to a quartic polynomial
$\cQ(T,U)$ in $X^2=T,\ X^3=U$ and $X^4=TU$. As we will see, for smaller
gauge groups, ${\cal C}$ takes a
different form and the duality group $SL(2,\ZZ)_T\times
SL(2,\ZZ)_U\times \ZZ_2^{T\leftrightarrow U}$ is broken down to a
subgroup (c.f., Table 1).

The correction $\Delta_{\F_T^4}$ represents a function of weights
$(w_T,w_U)=(4,-4)$ and $(w_{\ov T},w_{\ov U})=(0,0)$, respectively. Its
holomorphicity is spoiled by non--harmonic terms arising from massless
string states running in the loop. A holomorphic, covariant quantity
may
be obtained via an additional $T$--modulus insertion in the four--point
function, by considering
$$
\GG_{TTTTT}=
-{i\o 8} {(U-\ov U)^2\o (T-\ov T)^3}
\int d^2 \tau \sum_{(P_L,P_R)}P_L\ov P_R^5\
q^{\h |P_L|^2}\ov q^{\h |P_R|^2}\ {\bar E_4^2\o \ov \eta^{24}}\ .
$$
It is a non-trivial feature that this integral indeed yields a
holomorphic
covariant quantity:
\begin{eqnarray}
\GG_{TTTTT}&=&
\lf(\p_T+{4\o T-\ov T}\ri)\lf(\p_T+{2\o T-\ov T}\ri)
\p_T\lf(\p_T-{2\o T-\ov T}\ri)\lf(\p_T-{4\o T-\ov T}\ri)\GG\nonumber \\
&=&\ {E_6(T)E^2_4(U)  \o [J(T)-J(U)]\eta^{24}(U)}\\
&=&\ \sum_{(k,l)>0} k ^5\,g(k l)\,
{{q_T}^{k}{q_U}^{l}\over 1-{q_T}^{k}{q_U}^{l}}
\ \equiv\ (\del_T)^5 \GG\ \ .\nonumber
\label{NICE}
\end{eqnarray}
This is similar for the other couplings, for example we find
\begin{equation}
\GG_{TTTUU}\
=\ -{1\o 2\pi i} {J'(T)\o J(T)-J(U)}-{1\o 2\pi i}
\p_T\ln\Psi_0(T,U)\ \equiv\ (\del_T)^3(\del_U)^2 \GG\ ,
\label{link}
\end{equation}
where
\begin{eqnarray}
\Psi_0(T,U)&=&
%e^{2\pi i T}
q_T
\prod_{(k,l)>0}\lf(1-
%e^{2\pi i(kT+lU)}
{q_T}^k {q_U}^l
\ri)^{
d(kl)}\ . 
\label{PSI}
\end{eqnarray}
$\Psi_0$ stays finite everywhere in the moduli space, i.e.,
$d(-1)=0= d(0)$, and thus represents a cusp form on
moduli space. The exponents are generated by
\begin{eqnarray}
\sum_{n>0} d(n)q^n\ &:=& \ {1\o 12}
{E_2^2E_4^2\o \eta^{24}}+{1\o 3}{E_2E_4E_6\o \eta^{24}}
-{5\o 24}{E_6^2\o \eta^{24}}-{5\o 24}{E_4^3\o \eta^{24}}.
%=-123120q-10713600q^2-371498400q^3-\ldots
\ 
\label{dcoeff}
\end{eqnarray}
Note that $\GG_{TTTUU}$ (and thus $\Psi_0$)  is not modular, in
contrast to $\GG_{TTTTT}$   (this is familiar from the $4d$ coupling
$\FF_{TTU}$, which also is not modular).\footnote{The appearance of the
non-modular function $E_2$ in  (\ref{dcoeff}) implies that the theorem of
Borcherds \cite{borch} concerning the modular weight of product functions
$\Psi$ does not apply here (at least not directly).}

The relevant physical feature of the four-point couplings
is that they have a logarithmic singularity of the form
\begin{eqnarray}
\taueff(T,U)\ &=&\ \Coeff1{2\pi i}\ln[J(T)-J(U)] + \Coeff1{2\pi i}
\ln[{\rm cusp\ form}]\ .
\label{logj}
\end{eqnarray}
Similar to the analogous situation in four dimensions
\cite{dieter,enhancements},
the $J$-functions encode the gauge enhancements pertaining to the
compactification torus $T^2$: $SU(2)$ for $T=U$, $SU(2)\times
SU(2)$ at $T=U=i$ and $SU(3)$ at $T=U=\rho\equiv e^{2\pi i/3}$, and in
particular reflect the charge multiplicities of the states becoming
light near the singularities. Specifically, near the $SU(2)$ locus the
prepotential looks:
\begin{eqnarray}
\Li_5\big[q_T(q_U)^{-1}\big]\ &\sim\ & a^4\ln[\sqrt{\a'}a]\ ,\qquad\
a\equiv \coeff1{\sqrt{\a'}}(T-U)\ ,
\label{leading}
\end{eqnarray}
and similar for the other gauge groups. This is indeed the
expected behavior (c.f., (\ref{lntau}) of the one-loop  field theory
effective action\footnote{For a recent exposition including references, see
eg.\ ref.\ \cite{tsey}.} of $d=8$ $N=2$ SYM.

\subsection{Example B: $SO(8)^4$ Gauge Symmetry}

The general structure of the one--loop corrections  is similar to
(\ref{ellgen}). The elliptic genus encodes the information about the gauge
group. Since we  break the $E_8\times E_8$ gauge group down to
$SO(8)^4$ with  Wilson lines, the Narain lattice $Z_{(2,2)}$ will be
affected.  More precisely, with the (discrete) Wilson line background
\begin{eqnarray}
a_1^I&=&\h(0,0,0,0,0,0,0,0;1,1,1,1,1,1,1,1)\ ,\nonumber \\ \\
a_2^I&=&\h(0,0,0,0,1,1,1,1;0,0,0,0,1,1,1,1)\ ,\nonumber
\label{wls}
\end{eqnarray}
\noindent the internal part $Z_{(18,2)}$ of the partition function
reads \cite{gins}
\begin{eqnarray}\label{parti}
Z^{SO(8)^4}_{(18,2)}(q,\ov q)&=&{T_2\o \tau_2}\sum_{(\bf h,\bf g)}
\sum_{\cA_{(\bf h;\bf g)}} e^{-2\pi i T\det \cA_{(\bf h;\bf g)} }
\ e^{-{\pi T_2\o \tau_2 U_2}|({1\atop U})^t
\cA_{(\bf h;\bf g)} ({\tau \atop 1})|^2}\ \cC_{(\bf h;\bf g)}(q)\ ,
\nonumber \\
\end{eqnarray}
with:
\begin{equation}
\cA_{(\bf h;\bf g)}=\pmatrix{ 2n^1+h_1 & 2l^1+g_1\cr\noalign{\vskip2pt}
            2n^2+h_2 & 2l^2+g_2\cr}\ ,\ h_i,g_i=0,1\ .
\label{MATRIX}
\end{equation}
In addition, we introduce the functions
$\cC_0:=\cC_{(0,0;0,0)}$\ ,\
$\cC_1:=\cC_{(0,0;0,1)}=$\newline$\cC_{(0,0;1,0)}=\cC_{(0,0;1,1)}$
with
\begin{equation}
\cC_0(q)={E_4^2\o \eta^{24}}\ \ ,\ \
\cC_1(q)={1\o 64}{(G_4+G_2^2)^2\o\eta^{24}}:=\sum_{n\geq -1}c_1(n)q^n\
\label{EE}
\end{equation}
(the $\Gamma_0(2)$ modular forms $G_2$, $G_4$ are defined in Appendix
\appB). The remaining functions $\cC_2=\theta_2^8\theta_3^8/\eta^{24}
:=\sum_{n\geq 0}c_2(n)q^n,\  \cC_3=\theta_2^8\theta_4^8/\eta^{24}$
follow from modular invariance. Inspecting the partition function
(\ref{parti})
shows that the model under consideration is equivalent to a
$\ZZ_2\times \ZZ_2$ toroidal orbifold\footnote{Without torsion, i.e.
twisted sectors like e.g.  $(\theta,\theta\omega)$ do not appear
\cite{torsion}.}, with K\"ahler modulus  $\tilde T=4T$ and the (freely
acting) shift
$\th=\h(0,0,1,0)\ , \ \omega=\h(0,0,0,1)$  on $(P_L,P_R)\in  {\cal
N}_{2,2}$, accompanied with the shifts $\Theta=a_1\ ,\ \Omega=a_2$ in
the $SO(32)$ root lattice. Moreover, thanks to the relation
$\cC_1+\cC_2+\cC_3=\cC_0$,  the partition function (\ref{parti}) boils down to
the ``N=2 subsectors'' $i=1,2,3$ of a $\ZZ_4$--orbifold that  has
appeared in \cite{msi}

\begin{eqnarray}
Z_{(2,2)_i}(q,\ov q)&=&\nu_i\
\sum_{A_i}e^{2\pi i\tau(m_1n^1+m_2n^2)}
e^{-{\pi\tau_2 \o \tilde T_2U_2}|\tilde TUn^2+\tilde Tn^1-Um_1+m_2|^2}\
,
\label{maysti}
\end{eqnarray}
with Narain cosets $A_i$ and the volume factor
$\nu_i=vol(\cN_{{2,2}_i})=\{1,{1\o 4},{1 \o 4}\}$   (for example,
$A_1=\{m_j\in 2\ZZ, n^j\in \ZZ\}$ for the untwisted sector; for
further details see \cite{msi}).

We are now prepared to evaluate (\ref{ellgen}) for the corrections
\newline$\Delta_{a\b} \equiv \Delta_{\tr F^2_{SO(8)_\alpha}\!\!\tr
F^2_{SO(8)_\beta}}\!\!\!$, involving two gauge bosons from one
$SO(8)_\alpha$ and two from the same or another one, $SO(8)_\beta$
(traces are taken in the adjoint of $SO(8)$).  In addition, we also
consider $\Delta_\a\equiv \Delta_{\tr F^4_{SO(8)_\alpha}}\!$.  All of
these become a sum over three cosets that represent the orbifold
sectors

\begin{eqnarray}
\Delta_{\a\b}&=&\int{d^2\tau \o \tau_2} \sum_{i=1,2,3}
Z_{(2,2)_i}(q,\ov q)\
\lf(Q_{\alpha}^2-{1\o 4\pi\tau_2} \ri)\lf(Q_{\beta}^2-{1\o 4\pi\tau_2}
\ri)
\ \ov \cC_i(\ov q)\ (\a\not=\b)\nonumber \\
\Delta_{\a\a}&=&\int{d^2\tau \o \tau_2} \sum_{i=1,2,3}
Z_{(2,2)_i}(q,\ov q)\
\lf((Q_{\alpha}^2)^2-{1\o 2\pi \tau_2}Q_\alpha^2+
{1\o 16\pi^2\tau_2^2}\ri)\ \ov \cC_i(\ov q)\\\
\Delta_{\alpha}&=&\int{d^2\tau \o \tau_2} \sum_{i=1,2,3}
Z_{(2,2)_i}(q,\ov q)\
Q_\alpha^4\ \ov \cC_i(\ov q)\ . \nonumber 
\label{FFdec}
\end{eqnarray}
\noindent The charge operators act on the relevant $U(1)$ factors, in complete
analogy to considerations in refs.\ cite{ellis} (e.g.: 
$Q_{\alpha}^4\equiv E_4-9(8q {d\o dq}\ln \theta_\alpha-{1\o 3} E_2)^2$.)
They have the effect of derivatives
acting on the corresponding $SO(8)$ factors, contributing
$\cC_{D_4}:={1\o\sqrt 8}(G_4+G_2^2)^\h=\theta_3^2\theta_4^2$ in the
partition function (\ref{EE}). In total we get
\begin{eqnarray}
\Delta_{\a\b}
&=&\int{d^2\tau \o \tau_2} \ \  \lf\{-b_{\alpha\beta}+  Z_{(2,2)}(q,\ov
q,\tilde T,U)   x_{\a\b}\nonumber \\&&\ \ +
\sum_{i=1,2,3} Z_{(2,2)_i}(q,\ov q,\tilde T,U)
(z_{\a\b} \hat{\cB_i} +y_{\a\b})\ri\}
\\
\Delta_\a
&=&\h\int{d^2\tau \o \tau_2} \ \  \lf\{Z_{(2,2)}(q,\ov
q,\tilde T,U) -{2\o 3} \sum_{i=1,2,3}Z_{(2,2)_i}(q,\ov q,\tilde T,U)
\ri\}\nonumber
\label{FFFF}
\end{eqnarray} 
\noindent with  $x_{\alpha\alpha}=-4$, $y_{\alpha\alpha}=4$,
$z_{\alpha\alpha}={1\o 4}$ and $b_{\alpha\alpha}=4$, plus $x_{12}=2$,
$y_{12}=-4$, $z_{12}={1\o 4}$ and  $b_{12}=-2$.  Above, the functions
$\cB_i$ are defined by
\begin{eqnarray}
\cB_1&=&{1\o \eta^{24}}\cC_{D_4}^2\lf(q{d\o dq}\cC_{D_4}\ri)^2=
{1\o 36\cdot 64} {(G_4+G_2^2)^2\o \eta^{24}}[E_2-\h
(\th_3^4+\theta_4^4)]^2\nonumber \\
\cB_2&=&
{\theta_2^4\theta_3^4\o \eta^{24}} \lf(q{d\o
dq}\theta_2^2\theta_3^2\ri)^2= {1\o 36} {\theta_2^8\theta_3^8\o
\eta^{24}} [E_2+\h(\theta_2^4+\theta_3^4)]^2
\ , \\
\cB_3&=&
{\theta_2^4\theta_4^4\o \eta^{24}} \lf(q{d\o
dq}\theta_2^2\theta_4^2\ri)^2= {1\o 36}
{\theta_2^8\theta_4^8\o\eta^{24}} [E_2+ \h (\theta_4^4-\theta_2^4)]^2.
\nonumber
\label{chargeinsertion}
\end{eqnarray}
(For $\hat{\cB_i}$ we need to replace $E_2$ by $E_2-{3\o \pi\tau_2}$ in
(\ref{chargeinsertion}).) According to \cite{msi} the structure of the
integrands in (\ref{FFFF}) and all related couplings is
$
[Z_1(\tau) +Z_1(-{1\o 4 \tau})+Z_1(-{1\o 4 \tau-2})    ]\cB_1(\tau).
%\eqn\atkinz
$
When introducing the orbits for the non--degenerate orbit and enlarging
the
integration region to the upper half--plane,
we turn the integrands into a contribution of a single coset:
\begin{equation}
Z_1(-{1\o 2\tau})[\cB_2(2\tau)+\cB_1({\tau\o 2})+\cB_1({\tau\o 2}+\h)]\
\label{single}
\end{equation}
The following
remarkable identity\footnote{Analogous to the Atkin-Lehner cancellations
of ref.\ \cite{GM}.}
\begin{equation}
\cB_2(2\tau)+\cB_1({\tau\o 2})+\cB_1({\tau\o 2}+\h)=16
\label{atkin}
\end{equation}
has the effect that the non-trivial $q$-dependence
cancels out in the harmonic contributions.
We thus eventually arrive at
\begin{eqnarray}
\Delta_{\a\a}\ &=&-4\ln[T_2U_2|\eta(U)|^4]+4\lf[\ln|\eta(4T)|^4-
2\ln|\eta(2T)|^4\ri]\nonumber\\&&\ \ {\rm non\hyp harm.} \nonumber\\ \\
\Delta_{\a}+{1\o 8}\Delta_{\a\a}\ &=&\ 
-{1\o 2}\ln[T_2U_2|\eta(2T)|^4|\eta(U)|^4]+{\rm non\hyp harm.} \nonumber
\label{deltaa}
\end{eqnarray}
and moreover:
\begin{equation}
\Delta_{12}\ =\ 2\ln[T_2U_2|\eta(U)|^4]-\Big[2\ln|\eta(4T)|^4
-4\ln|\eta(2T)|^4\Big]+{\rm non\hyp harm.}
\label{ffff}
\end{equation}
\noindent (We do not display here the non--harmonic piece, which is related to
$\hat{E_2}$ and which can be easily computed along the lines of \cite{HM}.)
According to the choice (\ref{wls}), the
coupling of $SO(8)_1$ to the remaining $SO(8)$'s are obtained by
expanding around the other two cusps at $T=0,\h$, which gives:
\begin{eqnarray}
\Delta_{13} &=&\! 2\ln[T_2U_2|\eta(U)|^4]+\Big[
2\ln|\eta(T)|^4-2\ln|\eta(2T)|^4|+2\ln|\eta(4T)|^4\Big]
%-i\Coeff \pi2|T|^2\!+{\rm n.h.}
\!\nonumber \\ && \ \ +{\rm non\hyp harm.}\nonumber \\ \\
\Delta_{14} &=& 2\ln[T_2U_2|\eta(U)|^4]+\Big[
-2\ln|\eta(T)|^4 +4\ln|\eta(2T)|^4\Big]
%-i\Coeff \pi2|T|^2
\!+{\rm non\hyp harm.}
\nonumber 
\label{gggg}
\end{eqnarray} 
Note that the $T,U$ sectors effectively decouple, which is
a consequence of the cancellation (\ref{atkin}).
For sake of completeness, we have also computed the corrections
$\Delta_{F_T^4}$ etc.\ in (\ref{intFV}). Since for these the analog of
\ref{single} is
\begin{equation}
Z_1(-{1\o 2\tau})[2^{-4}\cC_2(2\tau)+\cC_1({\tau\o 2})+
\cC_1({\tau\o 2}+\h)]=0\ ,
\label{atkintwo}
\end{equation}
these couplings vanish identically, and the underlying prepotential is
trivial. The $U(1)^2$ gauge symmetry cannot be enhanced to a
non--Abelian group, in accordance with a statement given in
\cite{polchwitt}.

\noindent This may be also seen by looking at the one--loop
correction $\Delta_{{\cal R}^4}$ to the gravitational coupling
$\cR\wedge\cR\wedge\cR\wedge\cR$.
The techniques described above may be applied to the
integral $\Delta_{{\cal R}^4} = \int{d^2\tau\o \tau_2} [\sum\limits_{i=1,2,3}
Z_i f_i-360]$, where we have $f_1= \Phi_2+256$. With a similar identity
as (\ref{atkin}), namely $f_2(2\tau)+f_1({\tau \o 2})+ f_1({\tau \o
2}+\h)=720$, we derive $\Delta_{{\cal R}^4}=-360 \ln
T_2U_2|\eta(2T)|^4|\eta(U)|^4 +const.$, which is obviously
non--singular.
Besides, we calculated the harmonic part of the one--loop correction
to $\tr {\cal R}^2\tr F^2_{SO(8)}$: $\Delta_{{\cal R}^2 F_{\alpha}^2} = 
-\int{d^2\tau\o \tau_2} [\sum\limits_{i=1,2,3}
Z_i \cE_i-144]=144\ln T_2U_2|\eta(2T)|^4|\eta(U)|^4 +{\rm non\hyp harm.}$,
with $\cE_1=\eta^{-24}\theta_3^8\theta_4^8E_2[E_2-\h(\theta_3^4+\theta_4^4)]$.
There is again no mixing between $T$ and $U$ in the harmonic part,
thanks to $\cE_2(2\tau)+\cE_1({\tau \o 2})+ \cE_1({\tau \o
2}+\h)=288$. 

\section{$F$-Theory Approach}
\subsection{7--Brane Interactions}
\setcounter{equation}{0}

Dual to the heterotic string on $T^2$ is $F$-theory on elliptic $K3$
surfaces \cite{Fth}, and thus the aim would be to reproduce the
prepotentials $\GG$ from suitable $K3$'s. Such surfaces must be
elliptic fibrations over $\IP^1$ (here coordinatized by $z$), and
$F$-theory compactified on them are defined to be type IIB string
compactifications on the base $\IP^1$ augmented with extra 7--branes.
The positions of the 7--branes are precisely the locations $z=z^*_i$ in
the base where the elliptic fiber $\cE$ degenerates.

Every 7--brane is characterized by some charges $(p,q)$, and encircling
it in the $z$--plane leads to a monodromy
\begin{equation}
M_{(p,q)}\ =\ \pmatrix{1+pq & q^2 \cr -p^2 & 1-p q\cr}\ \in\ SL(2,\ZZ).
\label{monodr}
\end{equation}
The monodromy transformations act on the type IIB coupling $\tau_s(z)
\equiv \cp0+i \bp0$ as fractional linear transformations, and in the
canonical way on the $SL(2,\ZZ)$ doublets $(\cp2,\bp2)$ and
$(\cp6,\bp6)$ (here, $\bp p, \cp p$ denote the NS-NS and R-R $p$-forms
that couple to $(p-1)$ branes). On the other hand, the self-dual 4-form
$\cp 4$ remains invariant. In general, the 7--branes are mutually
non-local with respect to each other, which means that $p q'-p'
q\not=0$ for two given branes.

Effective interactions in $8d$ space-time are generated by
superimposing
world-volume actions, and also by integrating out exchanges between the
7--branes. This will in general be very subtle, though, at least
because
of the generic mutual non-locality of the 7--branes. We will circumvent
this problem by considering only certain couplings and by restricting
to a favorable sub-space of the theory, where the relevant branes are
effectively local. We also will consider only the parity-odd
(wedge-product) terms in the effective theory, as these are generated
by simple exchanges of $(\cp p,\bp p)$ multiplets between branes.
Specifically, we will consider world-volume couplings of the form \cite{MD}:
\begin{eqnarray}
%I_{7\hyp brane}\ &=\  (\cp0+i \bp0) \F\wedge \F\wedge \F\wedge \F\, +
%%\,(p\cp2+q\bp2)\wedge \F\wedge \F\wedge \F \cr
%&+\, \cp4 \wedge \F\wedge \F + (p\cp6-q\bp6)\wedge \F+
%\cp 8\cdot 1 +....\ .
I_{(0,1) D7\hyp brane}\ &=&\  \cp 0 \F\wedge \F\wedge \F\wedge \F\, +
\,\cp2\wedge \F\wedge \F\wedge \F \\  &&+\, \cp4 \wedge \F\wedge \F +
\cp6\wedge \F+
\cp 8\cdot 1 +....\ .\nonumber
\label{Iaction}
\end{eqnarray}
For $(p,q)$ branes we will have instead $\cp2\to q \cp2+p\bp 2$
etc.\footnote{We will actually consider certain non-local compounds of 7--branes,
whose exact world-volume couplings are not known. However, for our
purposes numerical coefficients are not important, and all we need is
to assume the existence of couplings of the indicated generic form.}

The antisymmetric tensor fields behave like scalar fields on the base
$\IP^1$, and thus have Greens functions as follows:\footnote
{In the following, we will often write only the holomorphic
pieces of Greens functions, dropping the anti-holomorphic 
and non-harmonic pieces.}
\begin{eqnarray}
G^{(p)}(z_1,z_2)\equiv\langle\,\cp p(z_1),\cp{8-p}(z_2)\,\rangle\ &=&\
\ln[z_1-z_2]\nonumber \\ &&+{\rm non\hyp singular\ in\ }(z_1,z_2)\ 
\label{Cgreensfct}
\end{eqnarray}
and similar for $\bp p$. For a single pair of local 7--branes,
this leads to the following term in the effective action
\cite{cb,MDML,DKPS}:
\begin{eqnarray}
I\ &\sim\ &\taueff(z_1,z_2)\,(\F_1-\F_2)^4 \ , \qquad \ {\rm where} \nonumber \\
\taueff(z_1,z_2)\hp\  &=&{1\over 2\pi i}\ln[z_1-z_2]+{\rm finite}\ .
\label{taufrombranes}
\end{eqnarray}

As we will see, it is easy to match this logarithmic singularity,
coming from tree-level exchange of massless antisymmetric tensor
fields, with the logarithmic singularity in the heterotic 1-loop
couplings. This boils essentially down to find the map between the
locations $z_{1,2}$ of the relevant branes, and the heterotic moduli
$(T,U)$. This is what we will discuss, as a warm-up, in the next
section for the $E_8\times E_8$ model (Example A).

However, an exact matching of the full $(T,U)$ dependence is more
challenging, since this also requires the non-singular terms in
(\ref{taufrombranes}) to coincide. The problem includes in particular to
find the non-singular terms in the Greens functions (\ref{Cgreensfct}),  which
reflect the global structure of the $z$--plane, ie.,  the presence and
monodromies of all the 7--branes. This is in general a hard problem,
since we do not know suitable Greens functions on $K3$. However, we
will be able to use insight gained from Example A and map the problem
to a simpler one, which is to some extent tractable. We will then apply
this to Example B (with $SO(8)^4$ symmetry), and finally reproduce some
of the heterotic one-loop corrections from $K3$ geometry.
\subsection{Example A: $E_8\times E_8$ Gauge Symmetry.}

Since we have unbroken $E_8\times E_8$ gauge symmetry, the $K3$ surface
necessarily has $E_8\times E_8$ singularities. The Weierstra\ss\ form
of such a surface has been presented in \cite{Fth} and looks\footnote{Various
aspects of this model have been investigated in refs.\ \cite{CCLM,PADM}}
\begin{equation}
\PK\ =\
y^2+x^3+z^5w^7+z^7w^5 + \a\,x\,z^4w^4+\b\,z^6w^6\ =\ 0\ .
\label{kthree}
\end{equation}
In the patch $w=1$, this represents an elliptic fibration over the
$z$--plane, with $E_8$ singularities at $z=0$ and $z=\infty$.
%
%These singularities do not get resolved under perturbations by the
%``middle polynomials'' $x\,z^4w^4$ and $z^6w^6$, which form a
%distinguished closed subsector of the theory. This subsector encodes
%the data of the two-torus on the heterotic side.
%
The exact dependence of the $K3$
moduli $\a,\b$ on the heterotic moduli $(T,U)$ was found in ref.\
\cite{CCLM} by indirect reasoning and is given by
\begin{equation}
(48\a)^3\ =\ -J(T)J(U)\ ,\qquad (864\b)^2\ =\
(J(T)-1728)(J(U)-1728)\ .
\label{abJrel}
\end{equation}
We show in Appendix B that this, as expected, coincides with the
mirror map, ie., the map from  $\a,\b$ to flat coordinates, which
are indeed the natural coordinates $T,U$ of the heterotic string moduli
space.

\ni
The IIB coupling $\tau_s$ varies over the $\IP^1$ base according to
\begin{equation}
J(\tau_s(z))\ =\ 4\a^3 z^{12} (\Delta_\cE(z,\a,\b))^{-1}\ ,
\label{ellJ}
\end{equation}
where the discriminant of the elliptic fiber $\cE$ is
\begin{eqnarray}
\Delta_\cE(z,\a,\b)\ &=& \ z^{10}(4 \a^3z^2+ 27(1+\b z+z^2)^2)\ \nonumber \\ &\equiv:\ &
z^{10}\prod_{i=1}^4(z-z^*_i(\a,\b))\ .
\label{ellDelta}
\end{eqnarray}

This means that four 7--branes are located at the zeros $z^*_i$ of the
discriminant where $\cE$ degenerates. Moreover, ten 7--branes are
localized at the $E_8$ singularity at $z=0$ and ten more at $z=\infty$,
totalling 24.
%
%It has been shown in \doubref\Joh\GZ\ that the $E_8$
%singularities can be made out of seven 7--branes of type $(1,0)$ plus

%two of type $(1,-1)$ plus one of type $(3,-1)$. From \monodr\ one
%easily computes \doubref\Joh\GZ\ the monodromy around each of the
%singularities to be given by  $ST\equiv \left({0 \,-1\atop1\
%1}\right)$, with $(ST)^3=-\bfone$.
%
It is known that the monodromy around each of the $E_8$ singularities
is given by  $ST\equiv \left({0 \,-1\atop1\ 1}\right)$, with
$(ST)^3=-\bfone$. Thus only finite order ($\ZZ_6$) branch cuts emanate
from the ``$E_8$--planes'' at $z=0, \infty$.

The remaining four 7--branes, located at $z=z^*_i(\a,\b)$, encode the
relevant data about the two-torus on the heterotic side, the various
open string trajectories between them reflecting the BPS winding states
on the $T^2$. By carefully tracing the locations of these branes, we
have verified that two of them have $(p,q)$ charges given by $(1,1)$
and the other two have $(0,1)$.\footnote{This non-invariant statement
depends  on the basis and choice of paths, but the only important point
here is that the branes are mutually non-local.} If two or more of
these four 7--branes collide, there may be or may be not an enhanced
gauge symmetry, depending on the mutual locality properties. Enhanced
gauge symmetries only occur if the $K3$ discriminant\footnote {This is
identical to the discriminant of the $SU(3)$ Seiberg-Witten curve
\cite{suthree}, the two non-local Argyres-Douglas points \cite{AD}  corresponding
to $SU(3)$ gauge enhancement etc. This is because  the $K3$ surface
(\ref{kthree} can be brought by rescalings to the form $z+{1\over z}+ (x^3+a
x+b)+ y^2=0$,  and this differs from the SW curve only by a quadratic
piece. SW curves will indeed reappear further below.
%The six codimension-1 SW
%monopole singularities correspond to the $SU(2)$ gauge enhancements in
%$F$ theory, the three local crossings (at $\b=0$, $\a^3=27$) to
%$SU(2)\times SU(2)$ gauge symmetry, and the two non-local
%Argyres-Douglas points \AD\ (at $\a=0$, $\b=\pm2$) to $SU(3)$ gauge
%enhancement.
}
\begin{equation}
\Delta_{K3}\ =\ (4\a^3+27(\b+2)^2)(4\a^3+27(\b-2)^2)
\label{kthreedet}
\end{equation}
vanishes. Via (3.6), this  precisely matches the known gauge
enhancement loci in the heterotic $(T,U)$ moduli space, ie.,
$\Delta_{K3}\sim(J(T)-J(U))^2$.

For the purpose of computing pieces of the effective action, we do not
really know how to handle mutually non-local 7--branes.  A related
problem is how to distribute the field strengths $\F_T,\F_U$ on the
7--branes; we have four 7--branes, each locally hosting a $U(1)$, but
in
total only a $U(1)\times U(1)$ gauge symmetry within the $T,U$
sub-sector. The requisite reduction of independent $U(1)$ factors is
supposedly due to global effects \cite{Fth,MDML}, but the precise
mechanism has not yet been well enough understood in order to be
helpful for our purposes.

However, we can make life more easy by considering a certain
distinguished sub-space of the moduli space. It is given by $\a=0$,
which amounts to e.g. $U=\rho\equiv e^{2\pi i/3}$, and from (3.7) to
constant type IIB coupling, $\tau_s(z)\equiv\rho$. In this limit, pairs
of $(1,1)$ and $(0,1)$ branes coincide at
\begin{equation}
z^*_{1,2}(J(T))\ =\ {1\over2}\big(-\b(T)\mp\sqrt{\b^2(T)-4}\big)\ ,
\ \ \ \ \b(T)\equiv 2\sqrt{1-J(T)/1728}\ ,
\label{ziloc}
\end{equation}
to form what one may call ``$H_0$--planes'' (associated with Kodaira
singularity type $II$; we follow here the nomenclature of ref.\ \cite{GSM}).
Due to the non-locality, there is no gauge enhancement on the
$H_0$--planes. Similar to the $E_8$--planes discussed above, no net
logarithmic branch cuts emanate from the locations $z^*_{1,2}$ of the
$H_0$--planes: there are only a finite order monodromies given by
$(ST)^{-1}$.  Thus, in the $\a=0$ subspace, there isn't any object with
logarithmic monodromy left, so we effectively deal with only (almost)
local objects.

It is now easy to recover the singular behavior of the
heterotic one-loop amplitudes at $U=\rho$. The only branes that
can possibly give rise to the singularity are the two $H_0$--planes.
Their world-volume couplings are identical, and it is easy to see
that if we consider only the coupling
$\taueff=\GG_{TTUU}$, then the relevant contributions come from
\begin{equation}
\cp4\wedge \F_T\wedge \F_U
\label{relevCFF}
\end{equation}
(which reflects the $K3$ intersection form, $\Omega_{TU} =
\delta_{TU}$). Thus the singularity of this coupling is carried by
$\cp4$ exchange between the $H_0$ branes, and we immediately obtain
\begin{eqnarray}
\taueff(T,U\equiv\rho)\hp\ &=&\ \Coeff1{2\pi i}
\ln[z_1^*-z_2^*] + {\rm finite}\nonumber \\ \\
 &=&\ \Coeff1{2\pi i}\ln[J(T)] + {\rm finite}\ .\nonumber
\label{tauJ}
\end{eqnarray}
Remembering that $J(\rho)=0$, this exactly matches the heterotic result
 (2.10). To obtain the finite term is however much more complicated and
will not be attempted here for Example A (but we will compute
an exact result for Example B later).

Let us instead pause and see what can be learned
by reinterpreting the result (3.12), which was obtained by tree-level
closed string $\cp4$ exchange, in terms of open strings.

\subsection{The Mirror Map as a Map Between Open and Closed String Sectors}

We focus on the logarithmic singularity at $J(T)=0$, which reflects
gauge symmetry enhancement from $U(1)^2$ to $SU(3)$.  Expanding around
$T=\rho$ by writing $a\equiv {T-\rho\over T-\rho^2}$, we have
$J(a)\sim a^3+\cO(a^6)$, which gives a contribution to $\taueff$ of
$\coeff3{2\pi i}\ln[a]$. The factor of $3$ reflects the multiplicity of
the gluons becoming massless at $a=0$. These gluons correspond to open
strings stretching between the $H_0$--planes. Indeed, since $M_{(0,1)}$
and $M_{(1,1)}$ do not commute, there is an ambiguity as to what kind
of ``charge'' one can attribute to an $H_0$--plane. In effect, many
types of strings can end on a $H_0$--plane, and in particular a
degenerate triplet of mutually non-local strings of charges $[0,1]$,
$[1,1]$ and $[1,0]$ (plus their negatives) can link the two
$H_0$--planes (their $[p,q]$ charges correspond to the roots of
$SU(3)$); see Fig.1. These strings becomes tensionless if the two
 $H_0$--planes collide (for $\b=\pm2$), which then form a single
``$H_2$--plane'',\footnote{Note that the Kodaira elliptic singularity type
of the $H_2$--plane is not $I_3$ but $IV$ \cite{GSM}. This is why we get
$SU(3)$ with four colliding, mutually non-local 7--branes instead, as
usual, with three local ones.} associated with monodromy $(ST)^{-2}$.

More explicitly, in terms of open strings the logarithmic singularity
in (\ref{tauJ}) can be viewed as arising from perturbative one-loop
contributions of the form

\begin{equation}
\taueff \ \sim\ \sum_i {Q_i}^4 \ln[m_i]\ ,
\label{lnsum}
\end{equation}
where the sum runs over all relevant individual states separately. The
mass of a single BPS string of type $[p,q]$, stretching between some
locations $z_1$ and $z_2$, is according to \cite{Sen,SenII} given by
$$
m_{p,q}\ =\ |w_{p,q}(z_1)-w_{p,q}(z_2)|\ ,
$$
\noindent with $w_{p,q} =\int  dw_{p,q}$, where the metric
on the $z$--plane is
\begin{equation}
dw_{p,q}(z)\ =\ {1\over \varpi_0}(p+\tau_s(z)\,q)\,\eta(\tau_s(z))^2
\prod_i(z-z_i^*)^{-1/12}\,dz\ .
\label{zmetric}
\end{equation}
\newpage
\begin{figure}

\centerline{\psfig{file=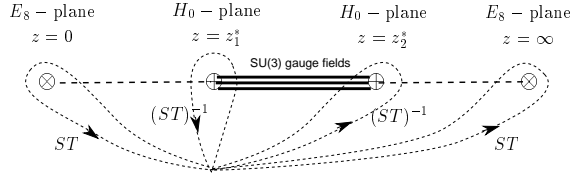,width=3.0in}}

\caption{\label{Hzero} {\small In the limit $\a=0$, pairs of non-local 7--branes
combine into $H_0$--planes with finite order monodromy. The planes are
connected by triplets of mutually non-local open strings. For $\b=\pm2$
the $H_0$--planes further merge into a single $H_2$--plane, the strings
between them then giving rise to massless charged $SU(3)$ gauge fields.
They reflect the intersecting vanishing cycles of an Argyres-Douglas
point.  }}
\end{figure}
Here, $z_i^*$ are the zeros of the elliptic discriminant $\Delta_\cE$
and $\varpi_0$ denotes the fundamental period of the $K3$
surface.\footnote{The prefactor $\varpi_0$ plays no r\^ole in the rigid
limit
discussed in \cite{Sen}, but must be there  in order to provide the correct
normalization of the BPS masses.}
In our situation with $\a=0$, where $\tau_s(z)\equiv\rho$ is constant,
the $\tau_s$-dependent terms are irrelevant and the integral can be
done explicitly. We so find that the mass of a string stretched between
the two $H_0$--planes (located at $z_{1,2}^*$ (\ref{ziloc})) is proportional
to
\begin{eqnarray}
\int_{z_1^*(j)}^{z_2^*(j)}&&z^{-5/6}
(z-z_1^*(j))^{-1/6}(z-z_2^*(j))^{-1/6}\,dz
\cr &=&
(i\sqrt{j(1-j)}-j )^{5/6} (-j)^{-1/12}\, \nonumber \\
&&\times{}_2F_1\big(5/6,5/6,5/3; 2j-2i\sqrt{j(1-j)}\big)\ ,\nonumber
\end{eqnarray}
\noindent where $j\equiv J(T)/1728$. This expression, via a non-trivial
hypergeometric identity, turns out to coincide with the $K3$ period
$f_2(J(T))$ defined in (B.6) in Appendix B. Dividing out
$\varpi_0$
(B.7) we see that the mass of the stretched string is given by a
certain
combination of the geometrical $K3$ periods $T,U$ (where
$U=\rho$):
$$
m_{p,q}\ =\ |(p+\rho\,q)\,a|\ ,
\qquad  a\equiv {T-\rho\over T-\rho^2}\ .
$$
The strings with $p,q$ charges $[1,0]$, $[1,1]$ and $[0,1]$ have thus
exactly the same, degenerate masses as the corresponding non-abelian
gauge bosons on the heterotic side. As $T\to\rho$, they indeed give the
leading contribution of ${3\over 2 \pi i}\ln[a]$ to (\ref{lnsum}).

In order to obtain the full expression (\ref{tauJ}) from (\ref{lnsum}), we would
need to take contributions of infinitely many open string states into
account. Their masses are given by distances in the $w$--plane and thus
by linear combinations of the $K3$ periods $(1,T,U,TU)$. On the other
hand, by simply considering distances in the $z$--plane (which governs
closed string $\cp p$ exchange), we had managed in the previous section
to sum up these infinitely many contributions in a single stroke
(signified by the appearance of a modular function, ie.,
$z_1^*-z_2^*\sim J(T)$).

In fact, it known (see e.g. \cite{DKPS}) that interactions between 7--branes
can be characterized by both open or closed string exchanges: open
string exchanges being dominant at short distances, while closed string
effects being important at large distances. In our example, this is
reflected by the expansion $J(a)\sim a^3+\cO(a^6)$, the short distance
variable $a$ being naturally adapted to open strings, while $J(a)$
provides the analytic continuation to the closed string regime.\footnote
{As we have already indicated, there are further corrections to ln$J$,
but for sake of easy reading we did not write them here.}

In other words, the map between open and closed string sectors is
provided by the period map between the $w$--plane and the $z$--plane,
and
thus is essentially nothing but the mirror map, $T\leftrightarrow
J(T)$.

\subsection{Constant Coupling $\tau_s$, and Reduction to Curves}
The above considerations can be canonically extended to the list of
examples in Table 1, including Example B with $SO(8)^4$ symmetry.  In
fact the $SO(8)^4$ model, introduced by Sen \cite{Sen}, was the first example
where the type IIB coupling $\tau_s$ is constant (and for this model
arbitrary) over the $z$--plane. Later it was found that there exist
also
other branches of the moduli space where the coupling is constant,
though frozen in value: $\tau_s=i$ (5 remaining moduli) or
$\tau_s=\rho$ (9 remaining moduli) \cite{KDSM}.\footnote{Various aspects of
theories
with constant coupling have been investigated in ref.\ \cite{GHZ}.}

We will restrict ourselves here to one-dimensional sub-spaces of the
moduli space of such theories,  given by the following one-parameter
families of elliptic $K3$'s:
\begin{eqnarray}
&(E_8,H_0):\ \ \ y^2 + x^3 + z^5(z-1)(z-z^*(\tau)) \ \ \ \,=\ 0 \nonumber \\
&(E_7,H_1):\ \ \ y^2 + x^3 + x z^3(z-1)(z-z^*(\tau)) \ \ =\ 0  \\
&(E_6,H_2):\ \ \ y^2 + x^3 + z^4(z-1)^2(z-z^*(\tau))^2 \ =\ 0 \nonumber \\
&(D_4,D_4):\ \ \ y^2 + x^3 + z^3(z-1)^3(z-z^*(\tau))^3 \,=\ 0\nonumber
\label{Kthrees}
\end{eqnarray}
Each of these models has two pairs of singularities in the $z$--plane
of
the indicated types, in generalization of Fig.1.  The first model is
equivalent to Example A ((\ref{kthree} restricted to $U=\rho$, i.e.,
$\a=0$), and the last one to our Example B (strictly speaking
restricted to $U=\rho$; however the $z$--plane geometry does not depend
at all on $U$, so our considerations will be valid for any
$U$.)

We have chosen to put
7--planes with symmetry $G_1=D_4,E_6,E_7,E_8$ at $z=0,\infty$, and
planes of type $G_2=D_4, H_2, H_1, H_0$, at $z=1,z^*$, respectively.
In fact, the Kodaira singularity types of these two sets are ``dual''
to each other, in that the monodromies of the $G_1$ planes and of the
$G_2$ planes are inverses of each other; they belong to $\ZZ_N$,
$N=2,3,4,6$, respectively.

In the one-dimensional moduli spaces, two interesting things can
happen. First, a $G_1$- and a $G_2$--plane can collide, to yield an
``$\hat E_8$'' singularity of the local form $y^2+x^3+z^6=0$. This
corresponds to the decompactification limit on the heterotic side,
$T\to i\infty$ \cite{PADM,FMW}. Second, similar to what we discussed
in the previous section, two $G_2$--planes can collide to produce a
7--plane associated with some extra non-abelian gauge symmetry $G_3$
(for $D_4$, where there is no further gauge enhancement, this also
corresponds to the decompactification limit). In other words, the
generic non-abelian gauge symmetry is $(G_1\times G_2)^2$, which can be
enhanced to $(G_1)^2\times G_3$. All this information is summarized in

{\vbox{{
$$
\vbox{\offinterlineskip\tabskip=0pt
\halign{\strut\vrule#
%%%%%%%%%%%%%%%%%%
&\hfil~$#$
&\vrule#
&~~$#$~~\hfil
&~~$#$~~\hfil
&~~$#$~~\hfil
&\vrule#
&~~$#$~~\hfil
&~~$#$~~\hfil
&~~$#$~~\hfil
&\vrule#
&~~$#$~~\hfil
&~~$#$~~\hfil
%&\vrule#
%&~~$#$~~\hfil
%&\vrule#
%&~~$#$~~\hfil
&\vrule#
\cr
%%%%%%%%%%%%%%%%%%
\noalign{\hrule}
& N
&&
G_1
&
{\rm Kod}
&
M
&&
G_2
&
{\rm Kod}
&
M
&&
G_3
&
{\rm Kod}
%&&
%{\rm  mod.\ subgroup}
%&&
%\tau_s
&
\cr
%%%%%%%%%%%%%%%%%%
\noalign{\hrule}
&
6
&&
E_8
&
II^*
&
ST
&&
H_0
&
II
&
(ST)^{-1}
&&
H_2
&
IV
%&&
%\Gamma^*
%&&
%\rho
&
\cr
%%%%%%%%%%%%%%%%%%
%\noalign{\hrule}
&
4
&&
E_7
&
III^*
&
S
&&
H_1
&
III
&
S^{-1}
&&
D_4
&
I_0^*
%&&
%\Gamma_0(2)
%&&
%i
&
\cr
%%%%%%%%%%%%%%%%%%
%\noalign{\hrule}
&
3
&&
E_6
&
IV^*
&
(ST)^2
&&
H_2
&
IV
&
(ST)^{-2}
&&
E_6
&
IV^*
%&&
%\Gamma_0(3)
%&&
%\rho
&
\cr
%%%%%%%%%%%%%%%%%%
%\noalign{\hrule}
&
2
&&
D_4
&
I_0^*
&
S^2
&&
D_4
&
I_0^*
&
S^2
&&
\hat E_8
&
-
%&&
%\Gamma_0(4)\cong\Gamma(2)
%&&
%{\rm any}
&
\cr
\noalign{\hrule}}
\hrule}$$
\vskip10pt}}}

{\vbox{{
$$
\vbox{\offinterlineskip\tabskip=0pt
\halign{\strut\vrule#
%%%%%%%%%%%%%%%%%%
&\hfil~$#$
&\vrule#
&~~$#$~~\hfil
&\vrule#
&~~$#$~~\hfil
&\vrule#
\cr
\noalign{\hrule}
& N
&&
{\rm  mod.\ subgroup}
&&
\tau_s
&
\cr
\noalign{\hrule}
&
6
&&
\Gamma^*
&&
\rho
&
\cr
&
4
&&
\Gamma_0(2)
&&
i
&
\cr
&
3
&&
\Gamma_0(3)
&&
\rho
&
\cr
&
2
&&
\Gamma_0(4)\cong\Gamma(2)
&&
{\rm any}
&
\cr
\noalign{\hrule}}
\hrule}$$
\vskip-10pt
\noindent{\bf Table 2:}
{\sl
Data of the two pairs of 7--planes in the four examples of Table 1:
$G_1$ and $G_2$ characterize the singularities associated with the
planes at $z=0,\infty$ and $z=1,z^*$, respectively (the $H_n$ planes
carry $A_n$ gauge symmetry). $G_3$ is the symmetry that appears if two
$G_2$--planes collide. ``Kod'' denotes the Kodaira singularity type,
and $M\in SL(2,\ZZs)$ the associated monodromy. We also indicate the
modular subgroup of which $z^*(\tau)$ is a modular function, and the
value of the constant type IIB string coupling $\tau_s$ (which
coincides with the value of $U$ in Table 1 up to an $SL(2,\ZZs)$
transformation; above, $\Ga^*$ denotes the modular group together with
some extra involution).
}
\vskip10pt}}}
By computations completely analogous to those in Appendix B, we
find directly from the $K3$'s in (\ref{Kthrees}) that the $z^*(\tau)$ satisfy
the Schwarzian differential equations
$$
{{z^*}'''\over {z^*}'}-{3\over2}({{z^*}''\over
{z^*}'})^2\ =\ -2Q({z^*})\,{{z^*}'}^2\ ,
$$
where
\begin{equation}
Q(z^*)\ =\ {N^2(z^*-1)^2+ 4(N-1)z^* \over 4N^2(z^*-1)^2 {z^*}^2}\ .
\label{Qdef}
\end{equation}
{}From these equations for $N=2,3,4,6$, we can easily infer what the
modular subgroups are of which $z^*$ is the Hauptmodul;\footnote{Explicitly, we find 
$z^*(\tau)=-{1\over A}[\eta(\tau)/\eta((6-N)\tau)]^{24/(5-N)}$
with $A=(16,27,64)$ for $N=2,3,4$ and 
$z^*(\tau)=(\sqrt {-j}+\sqrt{1-j})^{2}$ for $N=6$.} those subgroups
are listed in Table 2. It is reassuring that these precisely match the
modular subgroups that arise on the heterotic side by switching on the
corresponding Wilson lines (cf., Table 1). From (3.16) also follows
\cite{LY}
that the mirror map can be written as
\begin{equation}
\tau(z^*)\ =\ s(0,0,1-2/N; z^*)\ ,
\label{triangle}
\end{equation}
which supposedly matches the modulus $T$ on the heterotic side.
Above, $s(a,b,c)$ denotes the triangle function where the entries
$a,b,c$ denote the angles of a fundamental region. This means that
there are generically two cusps (corresponding to the
decompactification limits $z^*\to 0,\infty$) and one point with gauge
enhancement ($z^*\to 1$). However, for $N=2$ there are three cusps,
which reflects that two colliding $D_4$ singularities correspond to
decompactification. Indeed, for $N=2$ we have $z^*(\tau)=
\lambda(\tau)$
(the standard $\Gamma[2]$ modular function up to an $SL(2,\ZZ)$
transformation), which has three cusps only.

Note that since the coupling $\tau_s$ is constant,
the integrals over the open string metrics
\begin{eqnarray}
dw\ &=&\ dz\,\prod_{i=1}^{24}{1\over (z-z^*_i)^{1/12}}\ \nonumber \\  
&=& \ {dz\over z^{1-1/N}(z-1)^{1/N}(z-z^*)^{1/N}}
\label{ZNdiffs}
\end{eqnarray}
can be done explicitly. The two basic cycles give
the following hypergeometric functions:
\begin{eqnarray}
\varpi_0\ &=& \  (-1)^{-2/N} \pi\, {\rm csc}(\pi/N)\,
       {}_2F_1\big(1/N,1/N,1; z^* \big) \\
\varpi_1\ &=& {z^*}^{-1/N} (-1)^{-2/N} \pi\, {\rm csc}(\pi/N)\,
{}_2F_1\big(1/N,1/N,1; 1/z^* \big)\ ,\nonumber
\label{releper}
\end{eqnarray}
which coincide up to factors
with the relevant period integrals of the $K3$'s in (3.15). The flat coordinate is then alternatively given by
$\tau=\varpi_1/\varpi_0$.

An important observation is now that the holomorphic differentials
(\ref{ZNdiffs}) can also be viewed as abelian differentials $dw={1\over x}
{dz\over z}$ of the following genus $g=N-1$ Riemann surfaces:
\begin{equation}
\Sigma_{N-1}:\ \ \ x^N\ =\ z^{-1}(z-1)(z-z^*)\ .
\label{SWRS}
\end{equation}
These are nothing but $\ZZ_N$ symmetric variants \cite{otherZn} of the well-known
Seiberg-Witten curves for $SU(N)$ \cite{suthree}.  
Exactly these curves, as well as the differentials (\ref{ZNdiffs}), 
had in the past played a r\^ole in computing tree-level correlation 
functions in $\ZZ_N$ orbifolds \cite{hv,DFMS,EJSS}. 
That they appear now at this point is no accident,
since we have in fact a similar situation here: the $G_1$ and $G_2$
planes effectively figure as $\ZZ_N$ twist fields $\sigma_N$ and
$\bar\sigma_N$, respectively, with $\ZZ_N$ branch cuts running between
them in the $z$--plane.\footnote{Note that the
$z$--plane plays in this context the r\^ole of a tree-level world-sheet
with twist field insertions, and  not (for $N>2$) of a target space
$T^2/\ZZs_N$ (c.f., \cite{KDSM}).}

The $\ZZ_N$ monodromies are in particular felt by the $SL(2,\ZZ)$
doublets  $(\cp2,\bp2)$ and $(\cp6,\bp6)$, and correspond to the
monodromies inflicted on the scalar field $X(z)$ in a $\ZZ_N$ orbifold
compactification. We may thus view $X(z),\bar X(z)$ as given by
appropriate complex linear combinations of these NS-NS and R-R tensor
fields. On the other hand, the RR field $\cp 4$ is monodromy invariant
and so corresponds to a real scalar in the untwisted sector.
\begin{figure}
\centerline{\psfig{file=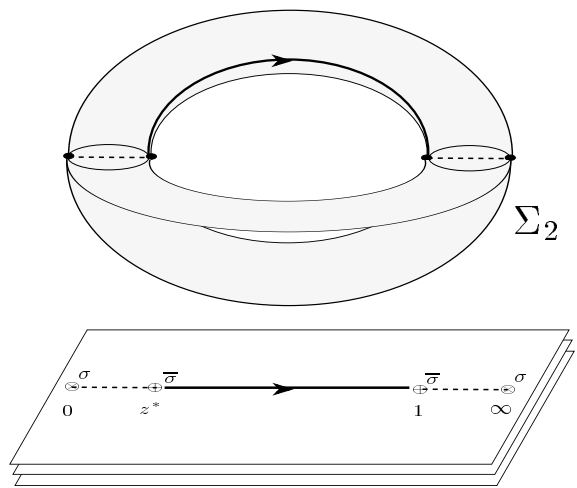,width=3.0in}}
\caption{\label{curve} {\small Lift of the $z$--plane to an $N$-fold cover, which is equivalent to the
$\ZZs_N$ symmetric  $SU(N)$ Seiberg-Witten curve $\Sigma_{N-1}$.  The
four 7--planes correspond to $\ZZs_N$ twist fields $\sigma$ and their
conjugates, $\bar\sigma$, and carry gauge symmetries $G_1$ and $G_2$,
respectively, as listed in Table 2. Shown is also an open string
trajectory that corresponds to a half--period on $\Sigma_{N-1}$.
  }}

\end{figure}
These observations then finally yield the following proposal for the
requisite Greens functions $G(z_1,z_2)$ in (3.3): they should
simply be given by the scalar Greens functions in the presence of
background twist fields $\sigma_N,\bar\sigma_N$ (with singularities
$X\to\sigma_N,\bar\sigma_N$ subtracted). 
According to \cite{hv,DFMS}, such Greens
functions can be described in terms of suitable covering spaces, and
those are precisely the curves of eq.\ (3.20) -- see Fig 2.
We thus expect that the Greens functions should simply be given by
appropriate Greens functions on the curves $\Sigma_{N-1}$.
That is, the fields that are twisted in the $z$--plane would
correspond to $\ZZ_N$--odd functions on $\Sigma_{N-1}$, while
single--valued fields (like $\cp 4$) would correspond to the
$\ZZ_N$--symmetric functions on $\Sigma_{N-1}$.

The Greens function of a scalar field on a Riemann surface is known
to be given by the logarithm of the prime form \cite{primef}
\begin{eqnarray}
G_\Sigma(z_1,z_2)\ &=&\ \ln\Big|{
\theta_\delta[\int_{z_1}^{z_2}\vec w\vert\Omega]
\over
\sqrt{\xi(z_1)}\sqrt{\xi(z_2)}
}\Big| \nonumber \\
&& \ -\pi\big[{\rm Im}\int_{z_1}^{z_2}\vec w]\!\cdot\!({\rm
Im}\Omega)^{-1}
\!\!\cdot\! \big[{\rm Im}\int_{z_1}^{z_2}\vec w]\ ,
\label{primeform}
\end{eqnarray}
where $\sqrt{\xi(z)}\equiv\sqrt{{\del\over\del z_i}(\theta_\delta[\vec
z\vert \Omega])\cdot w^i(z)}$ is a certain 1/2-differential that
cancels spurious zeros of the numerator theta-function (where $\delta$
denotes an arbitrary odd characteristic). Indeed, the only singularity
of the prime form is at coincident points, ie., $G_\Sigma(z_1\to
z_2)\sim\ln[{z_2-z_1\over\sqrt{dz_1} \sqrt{dz_2}}]+$ finite terms.  By
construction, the finite terms implement the requisite global
properties of the Greens functions. Of course, due to the high degree
of symmetry of our curves $\Sigma_{N-1}$, much of the information  in (3.21) is redundant in our examples for $N>2$ (e.g., the period
matrix $\Omega$ is proportional to the $A_{N-1}$ Cartan matrix and
depends only on one multiplicative parameter given by $\tau$); in
practice, one may wish to express the Greens functions more concisely
in terms of suitable, Weyl invariant Jacobi forms.\footnote{This might be
related to the considerations in \cite{kawaiII}.}

\subsection{Example B: $SO(8)^4$ Gauge Symmetry}
We now consider certain couplings in Example B. Though considerably
more complicated than Example A on the heterotic side, it is much
simpler than Example A on the $F$-theory side, because just a genus one
curve is involved (and not genus five). We will focus here on the
simplest couplings, namely on those which involve two different $SO(8)$
gauge fields (denoted by  $\Delta_{\alpha\beta}\equiv \Delta_{\tr
F^2_{SO(8)_\alpha}\tr F^2_{SO(8)_\beta}}$ in section 2.2). It is easy
to see that the only contribution to these couplings can come
from $\cp 4$ exchange between the two $D_4$--planes, each with
couplings
$$
\cp 4\wedge F_{SO(8)_\a}\wedge F_{SO(8)_\a}\ .
$$
As mentioned above, there is no gauge enhancement if two of such branes
collide. Rather, the singularities of pairwise collisions of the form
$\ln[z^*]$, $\ln[1-z^*]$ and $\ln[1/z^*]$ correspond to
decompactification limits on the heterotic side. Indeed, we know from
the previous section that $z^*=\lambda(\tau)$ (the $\Gamma(2)$ modular
invariant), and $\lambda=0,1,\infty$ are the three cusp points.

Thus all what we need is the Greens function that is appropriate to
$\cp 4$. Since this is an untwisted field, we need to consider a
$\ZZ_2$ even function on the torus $\Sigma_1$. Up to an overall factor 
of two \cite{hv}, this coincides with the ordinary Greens function on the torus:
$$
G(z_\a,z_\b)\ =\ \ln\left|{
\theta_1[\int_{z_\a}^{z_\b} w\vert \tau]
\over\theta_1'[\tau]/2\pi
}\right| - \pi 
{{\rm Im}^2(\int_{z_\a}^{z_\b} w)\over {\rm Im}(\tau) }
\ .
$$
The integrals are trivial since they just give the half-periods, 
\begin{eqnarray}
\theta_1[{\scriptstyle \int_{z^*}^0 w}]\ &=&\ \theta_1[\shalf|\tau]\ =
\ \theta_2 \nonumber \\
\theta_1[{\scriptstyle \int_{z^*}^\infty w}]\ &=&\
\theta_1[\shalf(1+\tau)|\tau]\ =
\ e^{-i\pi\tau/4}\theta_3 \nonumber \\
\theta_1[{\scriptstyle \int_{z^*}^1 w}]\ &=&\ \theta_1[\shalf
\tau|\tau]\ \ =\ ie^{-i\pi\tau/4}\theta_4\ , \nonumber
\end{eqnarray}
and therefore we have up to irrelevant constants:
\begin{eqnarray}
\Delta_{z^*,0}\hp\ &=&\ \ln\Big[{
\theta_2[\tau]\over\theta_1'[\tau]
}\Big]\nonumber \\
\Delta_{z^*,1}\hp\ &=&\ \ln\Big[{
\theta_3[\tau]\over\theta_1'[\tau]
}\Big]\\
\Delta_{z^*,\infty}\hp\ &=&\ \ln\Big[{
\theta_4[\tau]\over\theta_1'[\tau]
}\Big]\ .\nonumber
\label{Deleta}
\end{eqnarray}

Identifying $\tau$ with $2T$, and using the identities
$\theta_2/\theta_1'(\tau)=2\eta(2\tau)^2/\eta(\tau)^4$,
$\theta_3/\theta_1'(\tau)=\eta(\tau)^2/\eta(2\tau)^2/\eta(\tau/2)^2$
and $\theta_4/\theta_1'(\tau)=\eta(\tau/2)^2/\eta(\tau)^4$, we find
perfect agreement between the heterotic and the $F$-theory results ! So
indeed we have finally reproduced a piece of the effective action from
the relevant $K3$ surface in (3.15).

Observe that the sum of the three contributions (3.22) yields
$6\ln[\eta(2T)]$, which coincides (up to a factor) with the result
(2.22) that we got for the self-couplings, 
$\Delta_{\a}+\coeff18\Delta_{\a\a}$, $\a=1,...,4$. We
may interpret this as resulting from $\cp0-\cp8$ exchange between a
given brane and all the others, contributing  $\tr{F}^4+{1\o 8}(\tr
F^2)^2$ from the given brane and $1$ from the three others (c.f.,
(3.2)).
The same reasoning can be made for the
gravitational couplings $\Delta_{\cR^4}$ and $\Delta_{\cR^2\F^2}$, 
which indeed have the same functional form (see the end of
section (2.2)) and so can be obtained too by summing up
$C$-exchanges between all the 7-planes.
Note that the gravitational couplings (just like the $F^4$ couplings) arise
here in the geometric IIB string setup at tree level (and not at one-loop order, 
as one naively might have guessed.\footnote{Note that these parity-odd
gravitational $\cR^4$ terms have an index structure different to that of
the terms discussed in refs.\ \cite{MGMG}.})
Moreover recall that we have seen in section (2.2) that
one-loop corrections of type $\Delta_{\F_T^4}$ do not arise on the
heterotic side, and this is reflected here in the geometry  by the
absence of any other 7--branes besides the $SO(8)$--planes.

\section{Conclusions}
\setcounter{equation}{0}

The successful derivation of heterotic one-loop couplings from
classical $K3$ geometry represents one of the first quantitative tests
of the basic heterotic--F theory duality in eight dimensions. It also
reinforces the hope that mirror symmetry  would correctly sum up all
relevant perturbative and non-perturbative open string
interactions, in terms of tree-level closed string geometry.

However, this is not yet that clearcut: for the model for which we have
the strongest results, namely the $SO(8)^4$ model, the $U$ dependence
and thus the type IIB coupling constant dependence factors out.  This
means that the test we have made was not sensitive to non-perturbative
effects. For the other three models, the type IIB
coupling is fixed to a non-zero value and the $U$-dependence does not
factor out. Therefore the successful reproduction of an $F^4$ coupling
for one of those models would definitely be a non-perturbative test.
Unfortunately, the Greens functions that we have proposed for these
models are technically much harder to come by at higher genus,  and we
decided to leave this to future investigations. Moreover, the situation
is much more complicated if we consider couplings of types other than
$\Delta_{{\F_{G}}^2{\F_{G'}}^2}$, for which only $\cp4$ exchange was
relevant. To fully reproduce more general couplings, we would need to
work much harder, at least because for such couplings also twisted
field exchange is relevant, and possibly higher than two point
correlators. Also, bulk contributions cannot a priori be ruled out.
Clearly, a more detailed study is necessary in order to have a final
answer to the question whether or not classical $F$-theory geometry
captures the relevant quantum effects for {\it all} types of $F^4$
couplings.

Another interesting point is the mathematical 
interpretation of the integer ``instanton''
coefficients $g$ of the five-point 
couplings (1.4).\footnote{Formally, these
couplings look like the canonical holomorphic couplings associated with
five-folds. It would be interesting to find out whether five-folds
exist that reproduce these couplings, and if so, whether this would
have any physical significance. A naive first guess would have been
a fibration of $K3$ over $\IP^3$ (in the limit of large $\IP^3$), but, 
as we have checked, this does not work.} 
In fact, a simple interpretation can be given for the expansion
coefficients $\tilde g$ of the complete prepotential where all 18 moduli
are switched on. Specifically, noting that the factor ${E_4}^2$ in (2.5) can be attributed to the $E_8\times E_8$ partition function,
we are lead to conclude that the complete, unique prepotential that
contains all moduli is (up to a quintic polynomial,
in a particular chamber) given by
\begin{eqnarray}
\GG(T,U,\vec V)\ = \ -{i\over (2\pi)^5}
\sum_{{(k,l,\vec r)>0\atop \vec r\in \Lambda_{E_8\times E_8}}}&
\tilde g(kl-\vec r^{\,2}/2)
\times \Li_5\big[e^{2\pi i(kT+l U+\vec r\cdot\vec V)}\big]\ ,
\label{totalG}
\end{eqnarray}
where $\vec V$ are the $E_8\times E_8$ Wilson lines and where $\tilde g$
are the expansion coefficients of
\begin{equation}
\eta(q)^{-24}\ \equiv\  {1\over q}\prod_{l\geq1} (1-q^l)^{-24}
\ =:\ \sum_{n\geq -1} \tilde g(n)q^n\ .
\label{thetageneratingfunction}
\end{equation}
The various prepotentials corresponding to different gauge groups 
(e.g. those of Table 1) can be obtained as 
specializations of (4.1), since they correspond to different expansions 
around special points in the moduli space;
in particular, if we switch off all the $E_8\times E_8$ Wilson lines, 
we recover the prepotential (2.3). That the $\eta$-function
appears here as a counting function for $K3$ is perhaps not too
surprising, in view of the fact  that it counts 1/2-BPS states
on $K3$ \cite{VW,BSV,YZ} 
(corresponding to singular rational curves that are holomorphic
in a given, hyperk\"ahler-compatible complex structure).
Indeed it is known that the 
only contributions to the heterotic one-loop corrections to $F^4$
are due to 1/2-BPS states \cite{BK}.

Clearly, we have
followed in this paper a pragmatic physicist's approach, but we are
convinced that ultimately there should exist a much more direct 
route leading from $K3$ surfaces to the prepotential (4.1); 
this might be related to the considerations of ref.\ \cite{Dmirror}. 
Understanding the map between open 
and closed string sectors more generally
in terms of a mirror map may also to be useful for the recently
discovered relationships between gravity and gauge theory.

\centerline{\bf Acknowledgements}
%%%%%%%%%%%%%%%%%%%%%%% %%%%%%%%%%%%%%%%%%%%%%%%%%%%%%%%

We would like to thank
L.\ Alvarez-Gaum\'e,
M. Bershadsky, J.--P. Deren-\newline dinger, B.~de Wit,
 S.\ Ferrara,
M.\ Green, E.\ Kiritsis, A.\ Klemm, P.\ Mayr,
G.\ Moore, D.\ Morrison, S. Theisen, A.\ Todorov, C.\ Vafa,
and especially M.\ Douglas for discussions.
We are grateful to the ITP at UCSB for
hospitality, where the main part of this project was done.
This research was supported in part by the National
Science Foundation under Grant No.\ PHY94-07194.

\bigskip

\centerline{\bf Appendix A}
\vskip0.075in
\renewcommand{\theequation}{A.\arabic{equation}}
\setcounter{equation}{0}
\centerline{\bf Modular functions for $\Ga_0(2)$ and generalized $\tau$--integrals}
\bigskip
\noindent{\it A.1 Ring of Modular Forms}
\bigskip

The fundamental region of $\Ga_0(2)$ has two cusp points at  $\tau=0$
and $\tau=i\infty$. The ring of modular forms consists of two
generators of weights $2$ and  $4$, respectively:
\begin{eqnarray}
G_2&=&\th_3^4+\th_4^4\ , \\
G_4&=&(\th_3^4+\th_4^4)^2-2\th_2^8\ ,  \nonumber \\
\end{eqnarray}
with the $\theta$--functions:
\begin{equation}
\matrix{
\th_2=\sum\limits_{n\in \ZZ} q^{\h(n+\h)^2}\ , &
\th_3=\sum\limits_{n\in \ZZ}q^{\h n^2}
\cr\noalign{\vskip3pt}
\th_4=\sum\limits_{n\in \ZZ} (-1)^n q^{\h n^2}\ .  &}
\end{equation}

The unique normalized Hauptmodul $\Phi_2(\tau)$
of $\Ga_0(2)$ is given by
\begin{equation}
\Phi_2(\tau)=\lf[{\eta(\tau)\o\eta(2\tau)}\ri]^{24}=
q^{-1}-24+276q-2048q^2+\ldots\ .
\label{hauptmodul}
\end{equation}
To derive (2.19), we have used the following identities:
\begin{eqnarray}
q{d\o dq}G_2&=&{1\o 6}E_2G_2+{1\o 24}G_2^2-{1\o 8} G_4\\
q{d\o dq}G_4&=&{1\o 3}E_2G_4+{1\o 12}G_2G_4-{1\o 4} G_2^3\ .\nonumber
\label{derivatives}
\end{eqnarray}
In addition, note
\begin{eqnarray}
\cC_{D_4}^3 q{d\o dq}\lf(q{d\o dq}\cC_{D_4}\ri)&=&{1\o 8 \cdot 36}{1\o
64}
(G_4+G_2^2)^2\nonumber \\
&&(G_2^2-12E_2G_2+12E_2^2+3G_4-4E_4)\ ,
\end{eqnarray}
as well as the relation between $(F^2_{\alpha})^2$-- and
$F_{\alpha}^2F_{\beta}^2$--charge insertions
\begin{equation}
3\cC_{D_4}^2 \lf(q {d\o dq} \cC_{D_4} \ri)^2=2
\cC_{D_4}^3 q {d\o dq} \lf(q{d\o dq}\cC_{D_4}\ri) +{1\o 4}\eta^{24}\ .
\label{relation}
\end{equation}

\noindent{\it A.2. Generalized $\tau$--integrals}\newline
\vskip0.05in

In section 2.2 we had to deal with $\tau$--integrals whose 
integrands are certain sums over Narain cosets as they appear from orbifold
shifts; see e.g. the decomposition in (2.14)
for a $\ZZ_2\times \ZZ_2$ shifted lattice. After reducing 
the Narain lattice sum to (2.16), we could use 
techniques that were developed in \cite{msi} to perform the integrations. 
This method can be applied in general.
Let us demonstrate this with the $\ZZ_2$ coset:
\begin{equation}
\cA_{(h;g)}=\pmatrix{ n^1+\h h& l^1+\h g\cr\noalign{\vskip2pt}
            n^2 & l^2\cr}\ ,\ h,g=0,1\ .
\label{MATRIXtwo}
\end{equation}
\noindent Apart from the integral over the untwisted sector, which is of the kind 
treated in \cite{HM} 
(with the modular function $\cC_0-\cC_1-\cC_2-\cC_3$),
we have to do the integral\footnote{These cosets appear in the 
$\ZZ_8$--orbifold calculation of \cite{msi}, where further details may be 
looked up.} 
\begin{equation}
\int{d^2\tau\o \tau_2}\sum_{i=1,2,3}\nu_i\sum_{A_i}
\ov {P_R}_i^{a-1}
q^{\h|{P_L}_i|^2}\ov q^{\h|{P_R}_i|^2}\ \cC_i(\tau)\ ,
\label{basicintegral}
\end{equation}
with $\nu_i=vol(\cN_{2,2_i})=\{1,\h,\h\}$,  
$A_1=\{m_1\in 2\ZZ, m_2,n^1,n^2\in \ZZ\}$ and 
$A_2=\{n_1\in 2\ZZ, m_1,m_2,n^2\in \ZZ\}$ etc.
To perform the integration we may choose the orbits as in ref.\ \cite{msi}.
Generically, in the non--degenerate orbit 
(and part of the degenerate orbit), the three sectors $i=1,2,3$ sum 
up in the following combination:
\begin{equation}
\sum_{(k,l)>0}
2^{2-a} c_1(2kl)\ \Li_a(q_T^kq_U^{2l})+2^{2-a}c_2({kl\o 2})\ 
\Li_a(q_T^{k/2}q_U^l)+hc.\ .
\label{itworks}
\end{equation}
E.g. for $a=5$ this combination gives, up to a quintic polynomial,
the prepotential for the model with $[E_7\times SU(2)]^2$ gauge group
(the second entry in Table 1).
In this case $\cC_1={1\o 32}\eta^{-24}(G_4+G_2^2)G_2^2$, 
$T\rightarrow 2T,\ U\rightarrow U/2$ and again,
$\cC_0=\cC_1+\cC_2+\cC_3$.
For $a=1$, the trivial  orbit provides in addition 
${\pi \o 3}[c_1(0)+c_2(0)+c_3(0)-24(c_1(-1)+c_2(-1)+c_3(-1))]{T_2\o 2}$, 
if $\cC_i$ contains no $E_2$ factor.
The remaining part of the degenerate orbit
may be easily calculated along ref.\ \cite{HM} with the orbits of ref.\ \cite{msi}. 
There are further non--harmonic contributions, if $\cC_i$ contain powers
of $E_2$. However, again those have the structure (A.8) 
(with coefficients $\tilde c_i$ and multiplied by powers $T_2,U_2$)
and may be determined along refs. 
\cite{HM,ko} with the orbits of ref.\ \cite{msi}.
\newline
\bigskip

\centerline{\bf Appendix B}
\vskip0.05in
\centerline{\bf Mirror Map and $K3$ Periods for Example A}
\bigskip 
\renewcommand{\theequation}{B.\arabic{equation}}\setcounter{equation}{0}
We like to determine the dependence of the $K3$ moduli
$\a,\b$ in (3.5) on the flat coordinates $s_i\equiv\{T,U\}$.
For this, we closely follow the method detailed in
\cite{LSW}, and introduce the period integrals
\begin{equation}
u_0\ =\ \int {q(s)\over \PK}(\wedge dx_a)^4 \ ,\qquad\
u_k\ =\ \int {q(s)\phi_k\over {\PK}^2}(\wedge dx_a)^4\ .
\end{equation}

Here $\phi_k$ form a basis of the chiral ring (generated by
the middle polynomials $x z^4 w^4$ and $z^6 w^6$), and $q(s)$
is a conformal rescaling factor to be determined. These periods
satisfy the Gau\ss-Manin differential equation
\begin{equation}
{\del^2\over \del s_i\del s_j}u_0\ =\ c_{ij}^k u_k + \Gamma_{ij}^k
{\del\over \del s_k} u_0\ .
\end{equation}

In general, the conformal factor $q(s)$ is determined by
requiring that (B.2)
has no piece proportional to $u_0$, while $\a(s), \b(s)$ are
determined by the vanishing of the connection~$\G$.
More concretely,  let us focus on the ${\del^2\over \del^2 s_1}$ piece,
and denote the denominators proportional to $(\PK)^{-\ell}$ on the RHS
of (B.2.)  by $\kappa_\ell$:
\begin{eqnarray}
\kappa_1\ &=&\ q''(s)\nonumber \\
\kappa_2\ &=&\ -2 q'(s)\PK'(s,x_a)-q(s)\PK''(s,x_a)\\
\kappa_3\ &=&\ 2 q(s)(\PK'(s,x_a))^2\ ,\nonumber 
\end{eqnarray}

where the prime denotes derivatives wrt.\ $s_1$. As usual, the order of
$\kappa_3$ as a polynomial in $x_a$ can be reduced by
expanding into ring elements modulo vanishing relations,
and subsequently integrating by parts. The vanishing of the
connection $\Gamma$ then corresponds to the vanishing of the terms
proportional to $1/(\PK)^2$, and this gives:
\begin{eqnarray}
0\ &=&\ 2\b'(s)q'(s)+q(s)\big(-12\a(s)\b(s)(\a'(s))^2
\nonumber \\
&&+ 36\b(s)(\b'(s))^2+108\b''(s)+
4\a(s)^3\b''(s)-27\b(s)^2\b''(s)\big)\nonumber \\
0\ &=& \ 2\a'(s)q'(s)+q(s)\big(-6\a(s)^2(a'(s))^2
\nonumber \\
&&+ 18\a(s)(\b'(s))^2+4 \a(s)^3a''(s)-27(\b(s)^2-4)\a''(s)\big).\nonumber
\end{eqnarray}

Furthermore, the vanishing of the piece proportional to $u_0$
corresponds to the vanishing of the terms proportional
to $1/(\PK)$, and this yields:
$$
%\eqalign{
0=4\big(4\a(s)^3-27\b(s)^2+108\big)\,q''(s)+
\big(3 \b'(s)^2-\a(s)\a'(s)^2\big)\,q(s)\ .
$$
Similar equations are obtained by considering other second order
derivatives on the LHS of eq. (B.2). One easily checks that the
differential equations are solved by $q(s)=-{1\over
\eta^2(T)\eta^2(U)}$
and (3.6), indeed exactly as proposed in ref.~\cite{CCLM}.
%$$
%q(s) =
%\left({J(T)^{1/3}(J(T)-1728)^{1/4}\over\sqrt{J'(T)}}\right)
%\left({J(U)^{1/3}(J(U)-1728)^{1/4}\over\sqrt{J'(U)}}\right)
% = -{1\over \eta^2(T)\eta^2(U)}
%\ .
%$$

Writing $\a,\b$ in terms of $J$-functions (3.6)
drastically simplifies the Picard-Fuchs equations associated with
the $K3$ surface (3.5). These are given by
\begin{eqnarray}
\cL_1\ &=&\
\a+48\a\b\del_\b+24 \a^2\del_\a+48\a^2\b\del_\a\del_\b+
4(4\a^3-27\b^2+108){\del_\a}^2\nonumber \\
\cL_2\ &=&\ -3-144\b\del_\b-72\a\del_\a-144\a\b\del_\a\del_b+
4(4\a^3-27\b^2+108){\del_\b}^2\ .\nonumber 
\end{eqnarray}
Then indeed, transforming to variables
\begin{eqnarray}
J(T(\a,\b))\ &=&  -8\,\left( 4\,{{\alpha }^3} +
     27\,\left( -4 + {{\beta }^2} \right)\right) \nonumber \\ && 
-8\,\left(
     {\sqrt{1728\,{{\alpha }^3} +
         {{\left( 4\,{{\alpha }^3} +
              27\,\left( -4 + {{\beta }^2} \right)  \right) }^2}}}\right)
       \nonumber \\ \\
J(U(\a,\b))\ &=&\  8\,\left( 108 - 4\,{{\alpha }^3} - 27\,{{\beta }^2}\right) \nonumber \\ && +
     8\,\left({\sqrt{1728\,{{\alpha }^3} +
         {{\left( 4\,{{\alpha }^3} +
              27\,\left( -4 + {{\beta }^2} \right)  \right) }^2}}}
      \right)\nonumber
\end{eqnarray}
and rescaling the solutions, the PF system separates and
has the following solutions:
\begin{equation}
\tilde\varpi_i \ =\
J(T)^{1/12}J(U)^{1/12}\left(\matrix{
f_1(J(T)) f_1(J(U))\cr
f_2(J(T)) f_1(J(U))\cr
f_1(J(T)) f_2(J(U))\cr
f_2(J(T)) f_2(J(U))\cr
}\right)\ , \qquad i=0,...,3\ ,
%\ \sim\ \left(\matrix{1\cr U\cr T\cr TU}\right)
\end{equation}
where
\begin{eqnarray}
f_1(J)\ &=&\ {}_2F_1(1/12,1/12,2/3,J/1728)\\ 
f_2(J)\ &=& (J/1728)^{1/3} {}_2F_1(5/12,5/12,4/3,J/1728)\ .\nonumber
\end{eqnarray}
Dividing out $\tilde\varpi_0$ these solutions give the geometrical
integral $K3$ periods $\varpi_i=\{1,T,U,TU\}(\a,\b)$ up to linear
combinations. Specifically, in terms of analytically continued
functions,  we find for the fundamental geometric period
\begin{eqnarray}
\varpi_0\ &=&\ {}_2F_1(1/12,5/12,1,1728/J(T))\times
{}_2F_1(1/12,5/12,1,1728/J(U))\
\nonumber\\ &=&\
E_4(T)^{1/4}E_4(U)^{1/4}\ ,
\end{eqnarray}
exactly as for the well-known $K3$ of type
$X_{12}[1,1,4,6]$ \cite{KLM,LY}.
Indeed the PF systems of the two $K3$'s can be transformed into each
other, reflecting the universality of the $T,U$ subsector of the
moduli space. That this must be so also follows from the structure
of the relevant toric polyhedra.

\end{document}